\documentclass[]{pasj02} 

\jyear{2024}
\Received{}
\Accepted{}

\begin{document} 

\title{Elemental abundances of 44 very metal-poor stars determined from Subaru/IRD near-infrared spectra}

\author{
Wako \textsc{Aoki}\altaffilmark{1,2}\orcid{0000-0002-8975-6829}
\email{aoki.wako@nao.ac.jp}
Timothy C. \textsc{Beers}\altaffilmark{3}\orcid{0000-0003-4573-6233}
\email{timothy.c.beers.5@nd.edu}
Satoshi \textsc{Honda}\altaffilmark{4}\orcid{0000-0001-6653-8741}
\email{honda@nhao.jp}
Tadafumi \textsc{Matsuno}\altaffilmark{5}\orcid{0000-0002-8077-4617}
\email{matsuno@astro.rug.nl}
Vinicius M. \textsc{Placco}\altaffilmark{6}\orcid{0000-0003-4479-1265}
\email{vinicius.placco@noirlab.edu}
Jinmi \textsc{Yoon}\altaffilmark{3,7}\orcid{0000-0002-4168-239X}
\email{jyoon@stsci.edu}
Masayuki   \textsc{Kuzuhara}\altaffilmark{1,8}\orcid{0000-0002-4677-9182}
\email{m.kuzuhara@nao.ac.jp}
Hiroki  \textsc{Harakawa}\altaffilmark{9}\orcid{0000-0000-0000-0000} 
\email{harakawa@naoj.org}
Teruyuki \textsc{Hirano}\altaffilmark{1,2,8}\orcid{0000-0003-3618-7535}
Takayuki   \textsc{Kotani}\altaffilmark{1,2,8}\orcid{0000-0001-6181-3142}
\email{t.kotani@nao.ac.jp}
Takashi   \textsc{Kurokawa}\altaffilmark{8,10}\orcid{}
\email{tkuro@cc.tuat.ac.jp}
Jun   \textsc{Nishikawa}\altaffilmark{1,2,8}\orcid{0000-0001-9326-8134}
\email{jun.nishikawa@nao.ac.jp}
Masashi   \textsc{Omiya}\altaffilmark{1,8}\orcid{0000-0002-5051-6027}
\email{omiya.masashi@nao.ac.jp}
Motohide  \textsc{Tamura}\altaffilmark{1,8,11}\orcid{0000-0002-6510-0681}
\email{motohide.tamura@nao.ac.jp}
Sébastien  \textsc{Vievard}\altaffilmark{8,9}\orcid{0000-0003-4018-2569}
\email{vievard@naoj.org}
}

\altaffiltext{1}{National Astronomical Observatory, 2-21-1 Osawa, Mitaka, Tokyo 181-8588, Japan }
\altaffiltext{2}{Astronomical Science Program, Graduate Institute for Advanced Studies, SOKENDAI, 2-21-1 Osawa, Mitaka,
Tokyo 181-8588, Japan}
\altaffiltext{3}{Department of Physics and Astronomy and JINA Center for the Evolution of the Elements, University of Notre Dame, Notre Dame, IN 46556, USA}
\altaffiltext{4}{Nishi-Harima Astronomical Observatory, Center for Astronomy,
University of Hyogo, 407-2 Nishigaichi, Sayo-cho, Sayo, Hyogo 679-5313, Japan}
\altaffiltext{5}{Astronomisches Rechen-Institut, Zentrum f{\"u}r Astronomie der Universit{\"a}t Heidelberg, M{\"o}nchhofstra{\ss}e 12–14, 69120 Heidelberg, Germany}
\altaffiltext{6}{NSF NOIRLab, 950 N. Cherry Ave., Tucson, AZ 85719, USA}
\altaffiltext{7}{Space Telescope Science Institute, 3700 San Martin Dr., Baltimore, MD 21218, USA}
\altaffiltext{8}{Astrobiology Center, 2-21-1 Osawa, Mitaka, Tokyo 181-8588, Japan}
\altaffiltext{9}{Subaru Telescope, 650 N. Aohoku Place, Hilo, HI 96720, USA}
\altaffiltext{10}{Institute of Engineering, Tokyo University of Agriculture and Technology, 2-24-16, Nakacho, Koganei, Tokyo, 184-8588, Japan}
\altaffiltext{11}{Department of Astronomy, Graduate School of Science, The University of Tokyo, 7-3-1 Hongo, Bunkyo-ku, Tokyo 113-0033, Japan}




\KeyWords{nuclear reactions, nucleosynthesis, abundances --- stars:abundances --- stars: Population II}  

\maketitle

\begin{abstract}
Abundances of five elements, Na, Mg, Al, Si, and Sr, are investigated
for 44 very metal-poor stars ($-4.0 < $ [Fe/H] $ < -1.5$) in the
Galactic halo system based on an Local Thermodinamic Equilibrium (LTE)
analysis of high-resolution near-infrared spectra obtained with the
Infrared Doppler instrument (IRD) on the Subaru Telescope.  Mg and Si
abundances are determined for all 44 stars. The Si abundances are
determined from up to 29 lines, which provide reliable abundance
ratios compared to previous results from a few optical lines. The Mg
and Si of these stars are over-abundant, relative to iron, and are
well-explained by chemical-evolution models. No significant scatter is
found in the abundance ratios of both elements with respect to iron,
except for a few outliers. The small scatter of the abundance ratios
of these elements provides constraints on the variations of stellar
and supernova's yields at very low metallicity.  Al abundances are
determined for 27 stars from near-infrared lines (e.g., 1312~nm),
which are expected to be less affected by non-LTE (NLTE) effects than
optical resonance lines. The average of the [Al/Fe] ratios is close to
the solar value, and no dependence on metallicity is found over $-3.0
< $ [Fe/H] $ < -2.0$. Na abundances are determined for 12 stars; they
exhibit Solar abundance ratios and no dependence on metallicity. The
Sr abundances determined from the Sr II triplet are significantly
higher than those from the optical resonance lines obtained by
previous studies for our sample. This discrepancy shows a clear
dependence on temperature and surface gravity, supporting models that
predict large NLTE effects on the near-infrared lines for metal-poor
red giants.

\end{abstract}


\section{Introduction}
Elemental abundances of very metal-poor (VMP:
[Fe/H]$<-2$)\footnote{[A/B] = $\log(N_{\rm A}/N_{\rm B}) -\log(N_{\rm
  A}/N_{\rm B})_{\odot}$, and $\log\epsilon_{\rm A} =\log(N_{\rm
  A}/N_{\rm H})+12$ for elements A and B.} and extremely metal-poor (EMP: [Fe/H]$<-3$) stars contain unique
information on the details of nucleosynthesis pathways in the first
generations of massive stars and supernova explosions that constrain
the ranges of their initial masses
\citep{McWilliam1995AJ,Ryan1996ApJ,Nomoto2013ARAA,Ishigaki2018ApJ}. The trend and scatter of
elemental abundances of VMP stars trace the chemical evolution in the
early stages of the Galaxy's formation.

Among a variety of elements, the abundance ratios of $\alpha$-elements
and iron ([$[\alpha$/Fe], e.g., [Mg/Fe]) have been used as an
indicator of the contribution of type Ia supernovae compared to
core-collapse supernovae. 
The [$\alpha$/Fe] ratios of
VMP stars could, however, reflect the yields of massive stars that
are ejected by core-collapse supernovae, which depend on the
stellar mass, explosion energy, and other factors related to
the explosion mechanisms, while contributions of type Ia supernovae are
not expected, due to their long timescales. However, observational results
exhibit very small scatter in [Mg/Fe] in VMP stars (e.g.,
\cite{Cayrel2004AA, Yong2013ApJ}), except for a few outliers (e.g.,
\cite{Norris2001ApJ, Aoki2002ApJL, Ivans2003ApJ}). This suggests that
the yields of early generations of massive stars are similar in
$\alpha$-elements and iron, and/or that the gas clouds from which
VMP stars have formed were already well-mixed, homogenizing the [$\alpha$/Fe]
ratios. Si is another key element, because it is
as abundant as Mg, and is related to both the static nucleosynthesis
during stellar evolution and explosive nucleosynthesis in core-collapse supernovae. However, the
abundance ratios of [Si/Fe] in VMP stars are relatively
uncertain, because spectral lines that are useful for abundance
measurements for such stars are limited in the optical range (see below for more
details). Further estimates of the scatter of the abundance ratios for
other $\alpha$ elements are clearly desirable.

The abundance ratios of elements with odd atomic numbers, e.g., Na and
Al, are sensitive to the surplus of neutrons during C-shell burning
in massive stars, and depend both on the metallicity and adopted level of
atmospheric overshooting
(e.g., \cite{Kobayashi2006ApJ, Tominaga2007ApJ}). The abundance ratios
of these elements in VMP stars are uncertain compared to Mg, because
there are fewer measurements for VMP stars, as the numbers of (optical) spectral lines
that can be used for abundance measurements are relatively
small. Large NLTE effects are expected for absorption-line
formation for these spectral lines (e.g., \cite{Lind2022AA}), which
presents challenges for comparisons of measured abundance ratios with predictions by 
chemical-evolution models.

Sr is a key element to study the neutron-capture processes in the
early stages of chemical evolution. Sr could be produced by many
processes and sites, including the (main) $r$-process, the weak $r$-process
(\cite{Wanajo2006NuPhA}, also known as the Lighter Elements
Primary Process, LEPP; \cite{Travaglio2004ApJ}), as well as the weak $s$-process in massive stars
(e.g., \cite{Kappeler2011RvMP}). Sr has two strong resonance lines
in the blue spectral range (407.8 and 421.5~nm), which are detectable even for
extremely metal-poor (EMP; [Fe/H] $< -3.0$) stars. However, these lines are too strong to
determine accurate abundances in stars with relatively high
metallicity ([Fe/H] $\gtrsim -2.0$) or with excesses of Sr. Sr abundances
are less certain than abundances of another key neutron-capture
element, Ba, which has weaker lines in the red spectral range in addition to the two strong resonance lines (455.4 and 493.4~nm). 

We have previously reported on Si abundances determined from high-resolution
spectra in the $Y$-, $J$- and $H$-bands obtained with the Subaru Telescope
Infrared Doppler instrument (IRD; \cite{Tamura2012SPIE, Kotani2018SPIE})
for six metal-poor stars \citep{Aoki2022PASJ},
demonstrating that reliable Si abundances can be determined using 10-30
spectral lines of Si {\small I} that are free from blending by other spectral features, opening a new window
to investigate the abundance trends and scatter for Si in VMP and EMP
stars.  \citet{Aoki2022PASJ} also determined Sr
abundances from the Sr {\small II} triplet lines for four stars among
the six metal-poor stars. Whereas these Sr {\small II} lines in the near-infrared (NIR) range are free from blending 
and have strengths suitable for abundance measurements, the results
exhibit some discrepancy from the Sr abundances determined from
optical lines. This discrepancy may arise from NLTE effects
for the NIR lines, as discussed by
\citet{Andrievsky2011AA} and \citet{Bergemann2012AA}. More systematic
studies of Sr abundances from NIR spectra covering a wider
range of stellar parameters are also required to investigate the impact 
of NLTE effects on Sr abundance measurements.  The
{\it Y}-, {\it J}-, and {\it H}-bands include some other useful metallic
lines that are detectable for VMP and EMP stats (e.g., Mg, Fe),
although \citet{Aoki2022PASJ} focused on the above two elements.

Here we report abundance analyses of Si and Sr, as well as Na, Mg,
Al, and Fe, for 44 metal-poor stars in the range of [Fe/H] from $-4.0$ to $-1.5$
observed with the Subaru/IRD, including four of the six stars studied
by \citet{Aoki2022PASJ}. The sample selection and the IRD observations
are reported in Section~\ref{sec:obs}. Section~\ref{sec:ana} provides
details of the abundance analyses and error estimates. The abundance
results for Mg, Si, Na, and Al, and their trends with Fe, are discussed in
Section~\ref{sec:res}. The Sr abundance results are compared with
literature values obtained from optical Sr lines in this section, in
which NLTE effects are discussed. The results are compared with
predictions of Galactic chemical-evolution models in
Section~\ref{sec:disc}.

\section{Sample and Observations}\label{sec:obs}

Bright metal-poor stars were selected from literature for the
observations with Subaru/IRD. The list of the objects is provided in
table~\ref{tab:obj}. They are metal-poor stars that have been
well-studied by previous works to determine elemental abundances from
optical spectra. We focus on red giants, in which Si lines are
  detectable even in stars with very low metallicity
  (\cite{Aoki2022PASJ}). The effective temperature ($T_{\rm eff}$) and
[Fe/H] range from 4200~K to 5250~K and from $-4.0$ to $-1.5$,
respectively (see next section).  The sample includes carbon-enhanced
metal-poor (CEMP) stars (e.g., CS~30314-067, CS~29502-092). However,
the excesses of carbon for these stars are not very significant, and
no strong features of carbon-bearing molecules appear in the NIR
spectra. Four of the six stars in \citet{Aoki2022PASJ} are included in
the sample. The other objects are HD~25329, a cool main-sequence
metal-poor star, and the CH star HD~201626, which could be affected by
mass transfer from a companion AGB star.

\begin{figure}
  \begin{center}
       \includegraphics[width=16cm]{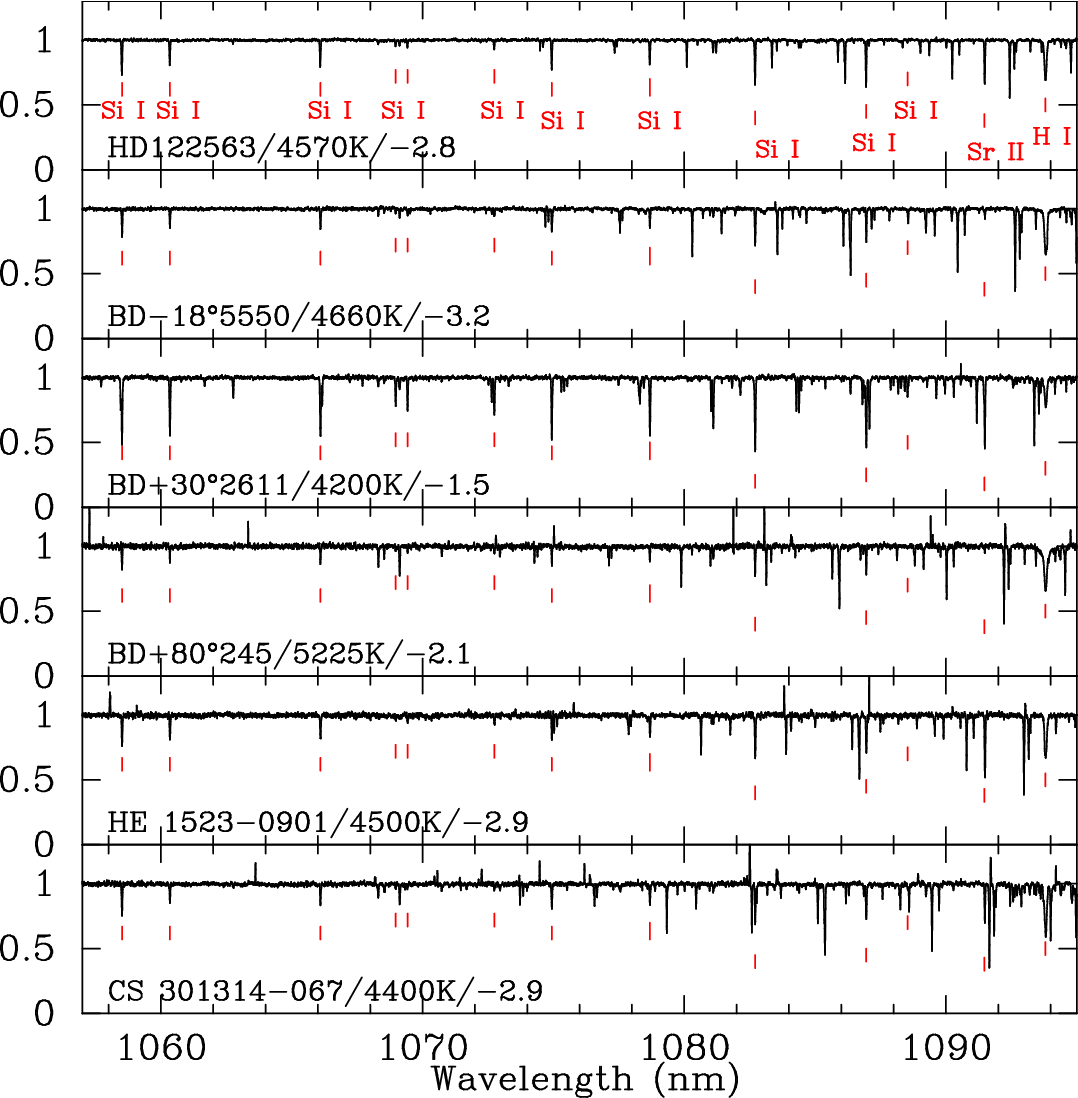}
\end{center}
\caption{Examples of near-infrared spectra obtained with
  Subaru/IRD. The star name, effective temperature, and metallicity are presented in each panel. The spectral
  lines of Si and Sr used for the abundance analysis, as well as a
  hydrogen line, are marked by red vertical bars. The CN absorption
  band at 1093--1095~nm is found for the carbon-enhanced star
  CS~30314-067. {Alt text: Six line graphs showing the normalized spectra of six stars. The x axis shows the wavelengths from 1057 to 1095~nm. The y axis shows the normalized flux density.}} \label{fig:sp}
\end{figure}

The NIR spectra were obtained with the Subaru/IRD on March 15, 2022
and November 8, 2022 (UT). The spectra cover the $Y$-, $J$- and
$H$-bands with spectral resolving power of $R\sim70,000$, with about
2.4~pixel sampling of the resolution element.
Examples of the spectra are shown in figure~\ref{fig:sp}.
The signal-to-noise (S/N) ratios per pixel are listed in table~\ref{tab:obj}, as estimated from photon counts at 1050~nm and 1600~nm.

We extracted the one-dimensional spectra of our targets from the raw IRD data using a custom-made pipeline, as described in \citet{Aoki2022PASJ}. The pipeline performs a series of data-reduction procedures (\cite{Kuzuhara2018SPIE}; Kuzuhara et al., in preparation). These include subtraction of bias and correlated read noise, removal of scattered light, correction for flat-fielding, tracing of stellar spectra, and wavelength calibration using Th-Ar emission spectra whose wavelengths were from the atlas of \citet{Kerber2008ApJS}. The pixel-to-wavelength relations were determined based on the Th-Ar lines and laser frequency comb spectra \citep{Hirano2020PASJ}.

Telluric absorption lines are identified by comparing the spectra of rapidly rotating early type stars, which exhibit only very broad and shallow absorption features, obtained in our observing program. Stellar spectral lines that are affected by telluric lines are excluded from the abundance analysis in the present work. 


 
\section{Abundance analysis}\label{sec:ana}
\subsection{Spectral line data and equivalent-width measurements}\label{sec:data}
Spectral line data of Si are given in \citet{Aoki2022PASJ}, who adopted
the data from VALD \citep{Kupka1999AAS} used by \citet{Fukue2021ApJ},
based on \citet{Kelleher2008}. Ten lines in the
$H$-band are supplemented from \citet{Afsar2016ApJ}, who take transition probabilities also from \citet{Kelleher2008} through NIST\footnote{National Institute of Standards and Technology:  https://physics.nist.gov/PhysRefData/ASD/lines\_form.html} for most of the lines. Fe lines in the
$Y$ and $J$-bands are also taken from the VALD data used by
\citet{Kondo2019ApJ}. Fe lines in the $H$-band are adopted from
\citet{Ruffoni2013ApJ}.  The line data for Na, Mg, and Al are taken
from \citet{Lind2022AA}, who adopt the data from \citet{Biemont1986PhyS} for Na, \citet{PehlivanRhodin2017A&A} for Mg, and \citet{Mendoza1995JPhB} for Al. The line data for the Sr triplet are taken
from \citet{Grevesse2015AA} as employed by \citet{Aoki2022PASJ}. The line
data, i.e., wavelengths, lower excitation potentials, and transition
probabilities ($\log gf$ values), used in the analysis are listed in table~\ref{tab:ew}. 

Equivalent widths are measured by fitting Gaussian profiles to the line profiles for our giant stars. The results are listed in table~\ref{tab:ew} with the above line data.  
Following the procedure of \citet{Aoki2022PASJ}, errors of the equivalent widths ($\sigma_{W}$) are estimated to be 0.2--0.9~pm, depending on the S/N ratios that range 50--300, by the formula of \citet{Norris2001ApJ} adopting $R=70,000$, $n_{\rm pix}=$10. The errors are slightly larger than 1.0~pm for LAMOST~J~0032+4107 and 2MASS~J~0643+5934 that have S/N ratios lower than 50. The abundance errors due to these equivalent width errors are discussed in the next subsection.


\subsection{Abundance determinations}

Abundances of Na, Mg, Al, Si, Fe, and Sr are determined from the equivalent widths by a standard
LTE analysis using model atmospheres from the ATLAS/NEWODF grid
\citep{Castelli2003IAUS} and the radiative transfer code, which is based on the same assumptions as the model atmosphere program of \citet{Tsuji1978AA}, following the procedure of \citet{Aoki2022PASJ} for the analysis of Si and Sr abundances.  The stellar parameters required for
abundance analysis based on model atmospheres are taken from the
literature, and listed in table~\ref{tab:param}. Effective temperatures
($T_{\rm eff}$) and surface gravities ($\log g$) in this table are used
with no modification in the analysis. This treatment is useful to compare the
abundance results from our NIR spectra with those of previous studies
from optical spectra, as for the Fe abundances described below. The [Fe/H] values are re-determined for most of the objects using the equivalent widths of optical Fe lines reported
in the literature (see below). The solar abundances reported by \citet{Asplund2009ARAA} are adopted to derive abundance ratios with respect to the solar values ([X/Fe]).

\begin{figure}
 \begin{center}
   \includegraphics[width=6.5cm]{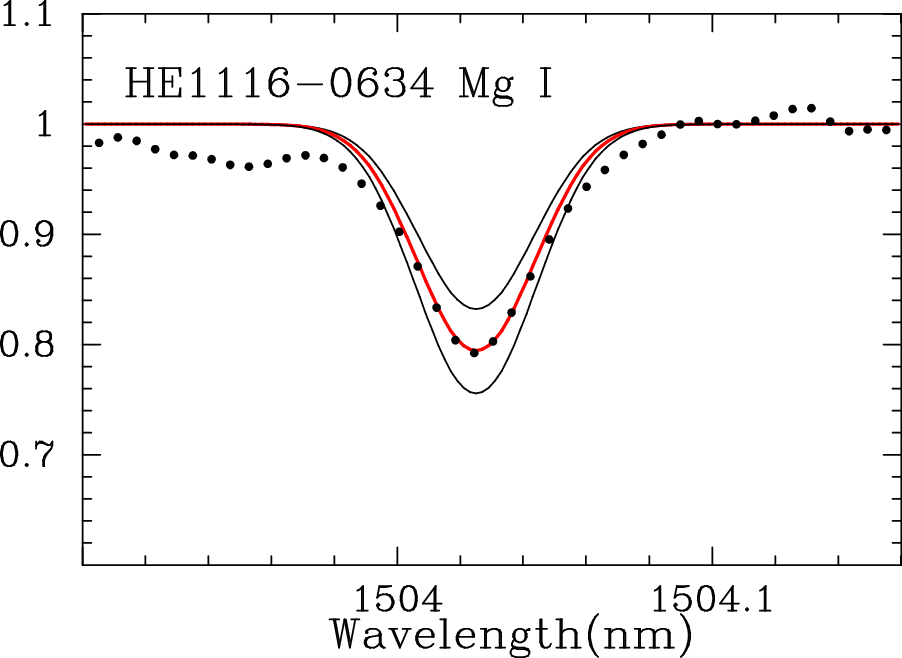}
   \includegraphics[width=6.5cm]{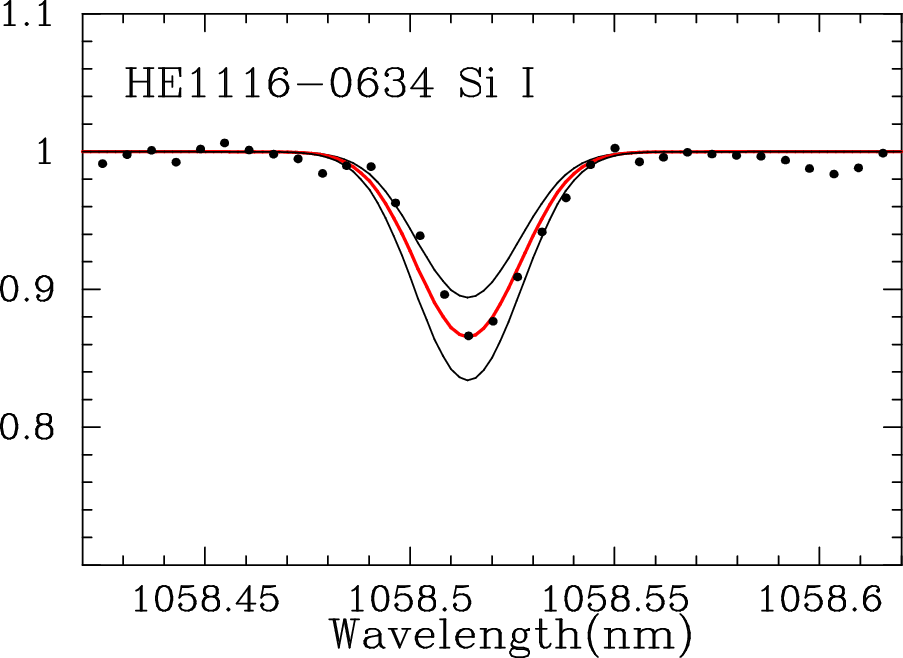}
   \includegraphics[width=6.5cm]{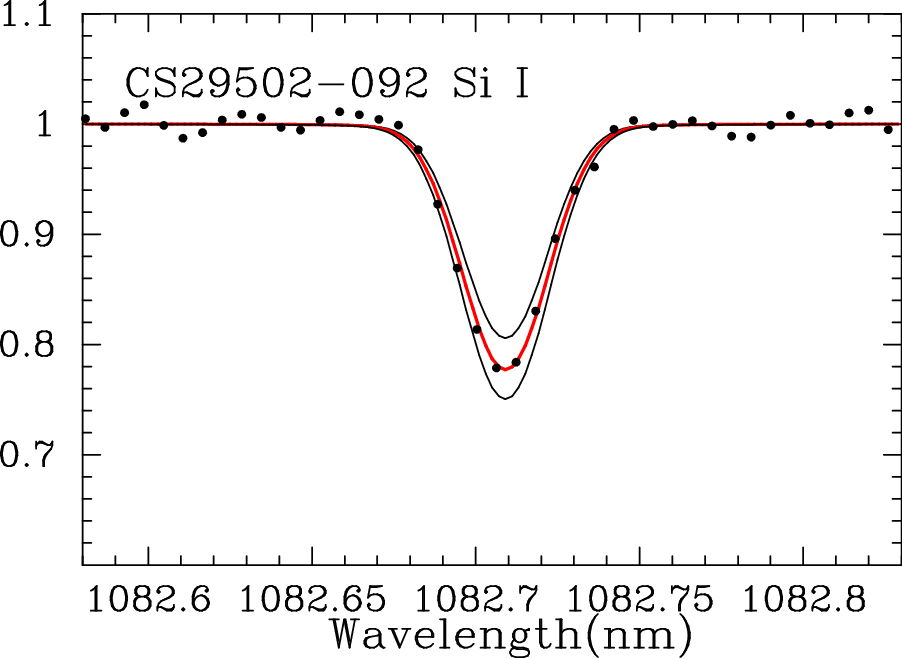}
   \includegraphics[width=6.5cm]{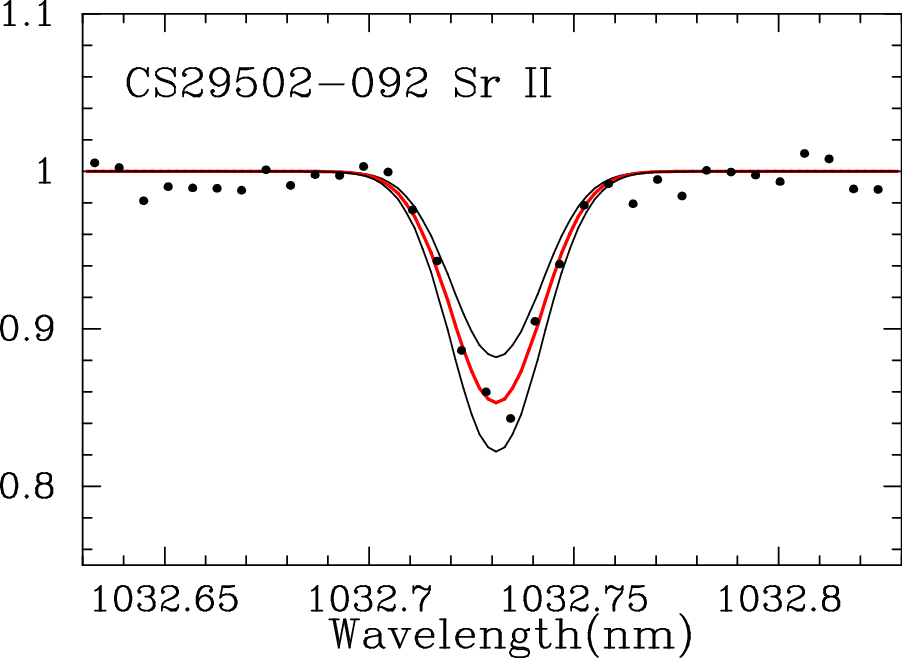}
 \end{center}
 \caption{Examples of specral lines used for the abundance analysis (dots). The object name and the species are presented in the panels. The synthetic spectra for the abundance determined from the line and those changed by 0.15~dex are shown by solid lines.  {Alt text: Four graphs showing observed spectra and synthetic ones. The x axis shows wavelengths and the y axis shows normalized flux.}}\label{fig:lines}

\end{figure}

Examples of spectral lines used for abundance measurements are
  shown in Figure~\ref{fig:lines} with synthetic spectra for the
  abundance determined from the line and those for 0.15~dex higher and
  lower abundances. We note that the abundances are determined by the
  analysis of equivalent widths. This figure is shown to demonstrate
  the data quality and the sensitivity of the spectral features to
  abundance changes.

Fe abundances are also measured from NIR lines for most of the objects
in our sample. The number of lines that can be used for the analysis
is larger than 10 for only 9 objects in our sample, stars that are relatively
metal rich. The average number of NIR lines available for the other stars is about 3. By contrast, many Fe lines are available in
the optical range, as studied by previous studies. 

To obtain more
reliable Fe abundances that are required to derive abundance ratios of
elements with respect to Fe ([X/Fe] values), we conducted a standard
analysis using the equivalent widths of optical lines reported by previous studies for
40 stars for which they are available. The references are given in table~\ref{tab:param}.
The table also lists the results ([Fe/H]$_{\rm opt}$), the number of lines used in the analysis ($N_{\rm opt}$), and the standard deviation of the Fe abundances derived from individual lines ($\sigma_{\rm opt}$). We found good agreement
between our results and previous studies ([Fe/H]$_{\rm ref}$): The
average and standard deviation of the differences between them for the
40 stars is 0.00~dex and 0.09~dex, respectively. We adopt the Fe
abundances derived from the optical lines by our analysis for the 40
stars, and those from the literature for the remaining 7 stars, as the
final results ([Fe/H]$_{\rm fin}$).

We list the Fe abundances derived from the NIR lines for 41 stars in table~\ref{tab:param} for reference. The results also agree
well with those from optical lines in general. The average and
standard deviations between the Fe abundances from the NIR lines and
the above final results are $-0.06$~dex and 0.11~dex, respectively. Four
objects exhibit discrepancies larger than 0.2~dex. The numbers of NIR lines
used for the analysis for these four stars are smaller than six, which
could result in relatively large uncertainties in Fe abundances derived from the
IRD spectra.

Si abundances are determined for all 44 stars in our sample. The
number of lines used in the analysis is from 14 to 29 for most
of these stars. The exceptions are BD+44$^{\circ}$493,
  LAMOST~J2217+2104 and LAMOST J~0032+4107, for which the number of
  available lines is 12 or less (table~\ref{tab:si}), because of their
  extremely low metallicity or limited S/N ratio. Even for these
  stars, five or more clean lines in the NIR range are used in the
  analysis, which improves the reliability of the Si abundances in
  these stars.


\citet{Afsar2016ApJ} report elemental abundances from optical and NIR
lines for two VMP stars, including HD~122563. The Si
abundance ratios ([Si/Fe]) of HD~122563 derived from the five lines in the $H$-band
used by their work and the present work are +0.46 and +0.50,
respectively, demonstrating very good agreement.


Mg abundances are also determined from 2 to 8 lines for all 44 stars. The number of lines depends on metallicity and data quality,
and is also significantly affected by the contamination from telluric
lines, which are present in the same spectral regions
as Mg. We note that the number of optical lines used to
determine Mg abundances is usually less than 10. Hence, measurements
from a few or several NIR lines could contribute to improving the
reliability of Mg abundance determinations.

\citet{Lind2022AA} reported the Na, Mg, and Al abundances determined
from optical and NIR lines for five stars,  including HD~122563, by LTE
and NLTE analyses. Five of the six Mg lines analyzed in the present
work are also used to determine the abundance in their work. The averages of the Mg abundances
($\log \epsilon$ values) determined from the five lines for HD~122563 by the
present work and the LTE analysis by \citet{Lind2022AA} are 5.27 and
5.25, respectively. The Mg abundance obtained from the same lines by
their NLTE analysis is 5.28, indicating that NLTE effects are very
small for this metal-poor red giant.

\citet{Afsar2016ApJ} also report the abundance results for Mg from
optical and NIR lines. The Mg abundance ratio ([Mg/Fe]) of HD~122563 from the
four lines used in their work and the present work are +0.46
and +0.45, respectively. 

Na abundances are derived from the Na I 1138~nm and 1140~nm lines. The
lines are weak, and severely affected by telluric lines in many stars. As a result,
the Na abundances are determined for only 12 stars. These lines are not well-studied previously, and no direct estimate of NLTE effects is reported. The estimates for other lines in the NIR range, which are summarized by \citet{Lind2022AA}, indicate that the effect is on the order of 0.1~dex for unsaturated lines. As the Na I lines detected in our spectra are weak, the NLTE effects for our results are not expected to be significant.

Al abundances are determined for 27 stars. The number of lines found
in the NIR range is at most six, and less than three in most
cases. Hence, the abundance results are not as reliable as for Si and
Mg. These lines are, however, still useful to determine Al abundances
for VMP stars because only two optical Al I lines
in the blue range are used in many cases of Al abundance studies for
VMP stars, and large NLTE effects are anticipated for these
lines (see next section).

The Al abundance ($\log \epsilon$ value) obtained for HD~122563 by
\citet{Lind2022AA} from the 2 Al lines that are also measured by the
present work is 3.75, which is 0.14~dex lower than our result
(3.89). The equivalent widths of the two lines used in their work are
smaller by about 40\% than ours, which is the likely reason for this
small difference in the abundance results. The reason for the discrepancy of equivalent widths between the two studies is not identified.

Sr abundances are determined for 40 stars in our sample. All three
lines of the Sr {\small II} triplet are used in most cases. The
remaining 4 stars include the two most metal-poor stars in our sample, BD+44$^{\circ}$493 and LAMOST~J2217+2104.  Note our spectrum of LAMOST~J0032+4107 has a relatively low S/N ratio. The other star is 2MASS J2016-0507, for which a very low Sr abundance ([Sr/Fe]$=-2.43$) is obtained by
\citet{Hansen2018ApJ} from the optical lines. The NIR Sr II lines are not detectable in stars with such low Sr abundances.


We estimate the abundance changes for changes of the stellar parameters,
$\Delta T_{\rm eff}=100$~K, $\Delta \log g=0.3$~dex,
$\Delta$ [Fe/H]$=0.3$~dex, and $\Delta v_{\rm turb}=0.5$~km~s$^{-1}$
for HD~221170 and 2MASS J0954+5246, which represent relatively metal-rich ([Fe/H]$\geq -2.5$) and more metal-poor ([Fe/H]$< -2.5$) stars in our sample. These parameter changes are typical uncertainties estimated for VMP stars in previous studies (e.g. \cite{Honda2004ApJ}).
The sensitivity of derived
abundances to the changes of stellar parameters is listed in
table~\ref{tab:error}.  The quadratic sum of the abundance changes
for the four parameters ($\sigma_{\log \epsilon}$) is also provided in the
table, along with the quadratic sum of the changes of
abundance ratios ([X/Fe]) that is presented as $\sigma_{\rm [X/Fe]}$. The $\sigma_{\rm [X/Fe]}$ is adopted as the error for the abundance  ratios of elements other than Fe. The error is
dominated by the changes of micro-turbulent velocity and effective
temperature (table~\ref{tab:error}).

The elemental abundances ($\log \epsilon$ and [X/Fe]), number of lines used in the analysis ($N$), standard deviation of abundances derived from individual lines ($\sigma$) and total errors including the errors due to uncertainties of atmospheric parameters (Error$_{\rm total}$) are listed in tables~\ref{tab:mg}, ~\ref{tab:si}, ~\ref{tab:na}, ~\ref{tab:al}, and  ~\ref{tab:sr} for Mg, Si, Na, Al, and Sr, respectively.
The $\sigma$ values would include the errors due to uncertainties of equivalent widths, which is 0.1~dex or smaller for the data quality of the current sample (Section~\ref{sec:data}). The uncertainties of spectral line data are smaller than 0.1~dex (see the detailed estimates for the analysis of Si lines in \cite{Aoki2022PASJ}). The random error of the abundance measurements is estimated by $\sigma N^{-1/2}$, which is included in the total error in the tables. The $\sigma$ of Si I ($\sigma_{\rm Si}$) is adopted in the
estimates for element X for which the number of lines available
in the analysis ($N_{\rm X}$) is small (i.e., the random error is $\sigma_{\rm Si}$/$N_{\rm X}^{1/2}$.

\section{Results}\label{sec:res}

\subsection{Mg and Si}

\begin{figure}
 \begin{center}
   \includegraphics[width=8cm]{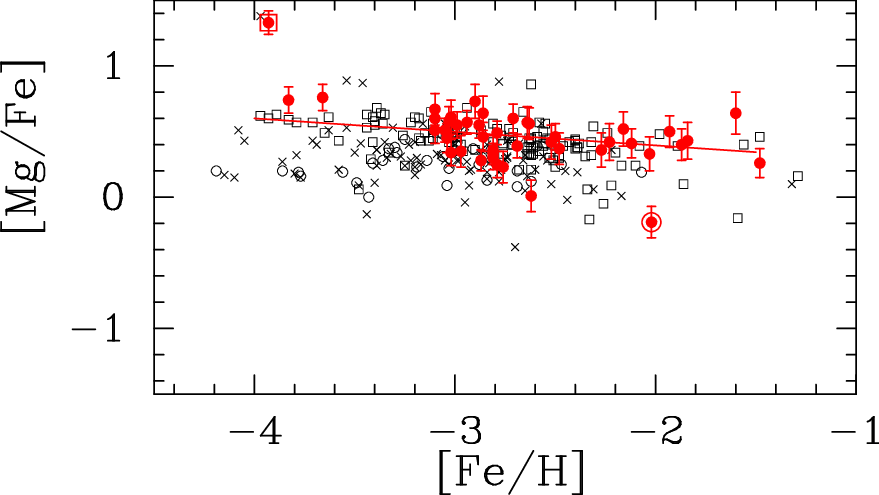}
   \includegraphics[width=8cm]{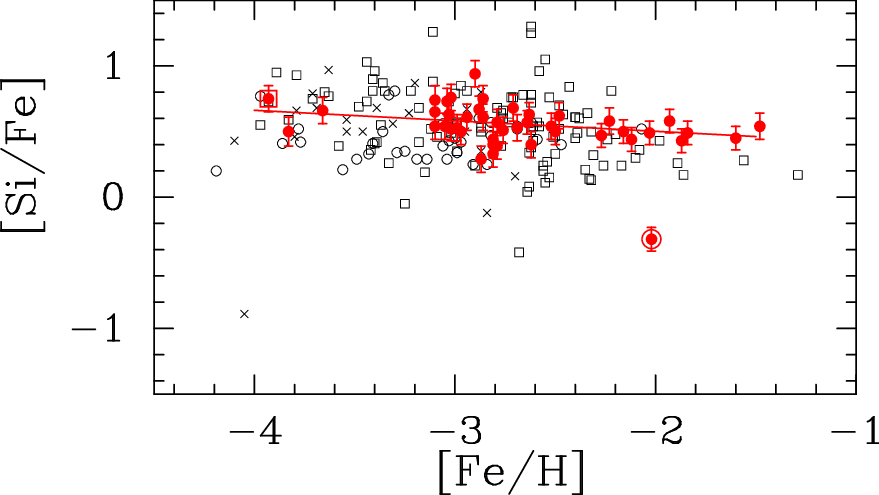}
   \includegraphics[width=8cm]{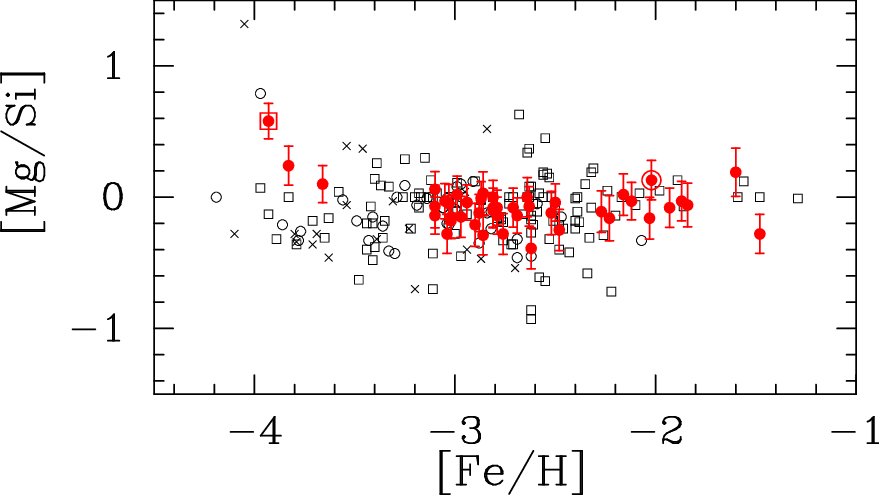}
 \end{center}
\caption{[Mg/Fe], [Si/Fe], and [Mg/Si], as a function of [Fe/H]. The
  results obtained by the present work are shown by filled
  circles. [Mg/Fe] and [Si/Fe] obtained by previous studies based on
  optical spectra are also shown in the corresponding panel by open
  circles: \citep{Cayrel2004AA}, crosses \citep{Yong2013ApJ}, and open
  squares \citep{Jacobson2015ApJ}. The $\alpha$-deficient star
  BD+80$^{\circ}$245 and the ultra metal-poor, carbon-enhanced star
  LAMOST~J2217+2104 are shown by large open circles and and open
  squares, respectively, over-plotted on the filled circles. The solid
  lines present the abundance trends obtained by least square fitting
  for the abundance ratios excluding the above two stars.
  {Alt text: Three graphs showing abundance ratios [Mg/Fe], [Si/Fe] and [Mg/Si]. The x axis shows [Fe/H] from $-4.5$ to $-1.0$ and the y axis shows the abundance ratios from $-1.5$ to +1.5.}}\label{fig:mgsi}

\end{figure}

The top two panels of figure~\ref{fig:mgsi} show [Mg/Fe] and [Si/Fe], as a function of [Fe/H], in the present work. For comparison, these panels also show the abundance ratios determined from optical lines by previous work for metal-poor stars, including substantial numbers of VMP and EMP stars \citep{Cayrel2004AA, Yong2013ApJ, Jacobson2015ApJ}. Mg abundance ratios obtained from NIR spectra in the present work agree quite well
with those obtained by the previous work. The Mg-enhanced ([Mg/Fe]$=+1.33$) star
at [Fe/H] $= -3.93$ is LAMOST~2217+2104 \citep{Aoki2018PASJ}, and the
Mg-deficient ([Mg/Fe]$=-0.19$) one at [Fe/H] $= -2.03$ is the well known
$\alpha$-deficient star BD+80$^{\circ}$245 (e.g.,
\cite{Ivans2003ApJ}). Excluding these two stars, the averages (
standard deviations) of [Mg/Fe] in the ranges $-3.5<$[Fe/H]$<-2.5$ and $-2.5<$[Fe/H]$<-1.5$ are 0.46 (0.15) and 0.43 (0.10), respectively (see table~\ref{tab:ave}).
A shallow, but statistically significant, slope of [Mg/Fe] is detected:
$\delta$[Mg/Fe]/$\delta$[Fe/H]$ = -0.10\pm$0.05.


The average of Si abundance ratios obtained from the NIR spectra agree well
with that from optical spectra by the previous work. The
scatter is, however, clearly smaller in the present results. We note
that the Mg-enhanced star LAMOST~J2217+2104 does not exhibit a remarkable
excess of Si compared to other stars. Excluding BD+80$^{\circ}$245, 
which also shows a significantly low Si abundance ratio, the averages (standard deviations) of [Si/Fe] in the ranges $-3.5<$[Fe/H]$<-2.5$ and $-2.5<$[Fe/H]$<-1.5$ are 0.58 (0.14) and 0.51 (0.05), respectively (see table~\ref{tab:ave}). The standard deviations are comparable with
the measurement error for [Si/Fe], which is typically 0.1~dex
(table~\ref{tab:si}). Our study for a large number of NIR Si lines
has revealed that the scatter of [Si/Fe] is as small as that of
[Mg/Fe] in VMP stars. Table~\ref{tab:ave} also lists the averages and standard deviations of [Mg/Fe] and [Si/Fe] determined by the three studies based on optical spectra \citep{Cayrel2004AA,Yong2013ApJ,Jacobson2015ApJ} shown in figure~\ref{fig:mgsi}. The larger scatter of [Si/Fe] than [Mg/Fe] found
by previous studies can be
accounted for by the larger errors in measurements based on a few Si lines in
the optical range.
A shallow slope of [Si/Fe] is also found:
$\delta$[Si/Fe]/$\delta$[Fe/H]$ = -0.08\pm$0.04. This result is compared with
chemical-evolution models in \S\ref{sec:disc}.

The bottom panel of figure~\ref{fig:mgsi} shows [Mg/Si], as a function of [Fe/H].   No significant scatter
and trends are found in [Mg/Si] over the metallicity range $-3.2 < $ [Fe/H] $ < -1.5$. Namely, Si
exhibits almost an identical abundance trend as Mg in this metallicity
range. We note that the [Mg/Si] of BD+80$^{\circ}$245 is
indistinguishable from other metal-poor stars. On the other hand, the
[Mg/Si] of LAMOST~J2217+2104 is higher than for other stars. The Si 
of this
star is over-abundant ([Si/Fe] = +0.75), as found by
\citet{Aoki2018PASJ}, but it is not as significant as found for Mg
([Mg/Fe] = +1.33). 

\subsection{Na and Al}

\begin{figure}
 \begin{center}
   \includegraphics[width=8cm]{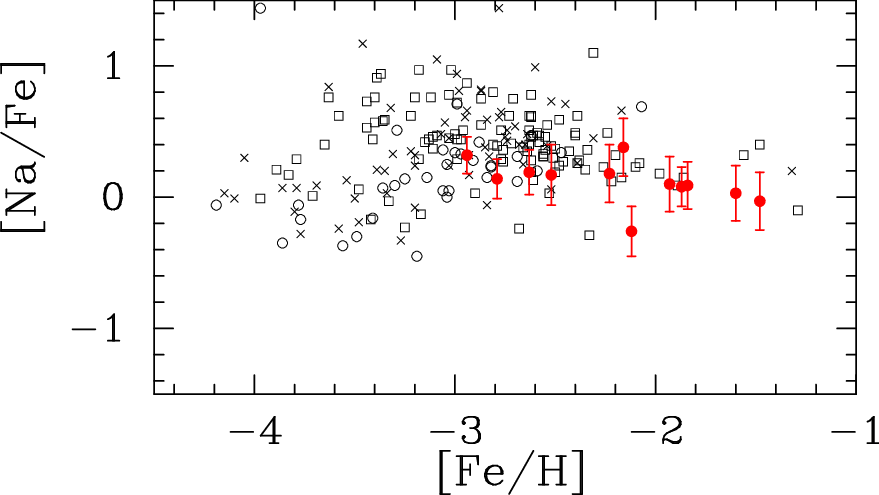}
   \includegraphics[width=8cm]{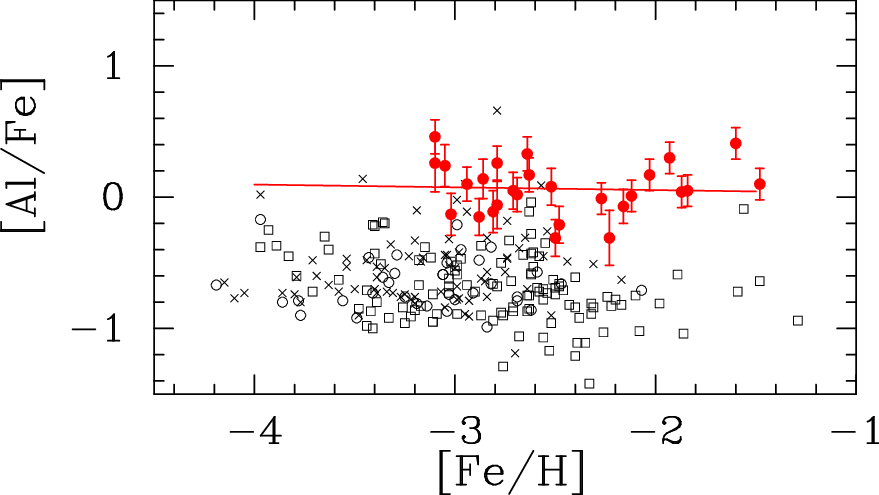}
 \end{center}
\caption{The same as figure~\ref{fig:mgsi}, but for [Na/Fe] and
  [Al/Fe]. The abundance trend is not shown for [Na/Fe]
  because the number of stars for which Na abundances are determined
  by the present work is too small. {Alt text: Two graphs showing abundance ratios [Na/Fe] and [Al/Fe]. The x axis shows [Fe/H] from $-4.5$ to $-1.0$ and the y axis shows the abundance ratios from $-1.5$ to +1.5.}}\label{fig:naal}
\end{figure}

Figure~\ref{fig:naal} shows [Na/Fe] and [Al/Fe], as a function of
[Fe/H]. The Na abundances are determined for only 10 stars in our
study. The average of [Na/Fe] is +0.17. This is lower than the results
of previous studies for this metallicity range ([Fe/H]$>-3$), which
mostly rely on the Na I D lines. Whereas the NLTE corrections for NIR lines is 0.1~dex level (\cite{Lind2022AA}), those for strong Na I
D lines could be as large as $-0.5$~dex for very metal-poor stars, as reported by, e.g.,
\citet{Andrievsky2007AA} and \citet{Lind2011AA}, as well as \citet{Lind2022AA}, which could be the
cause of the discrepancy between our results and previous ones.

\begin{figure}
 \begin{center}
   \includegraphics[width=6cm]{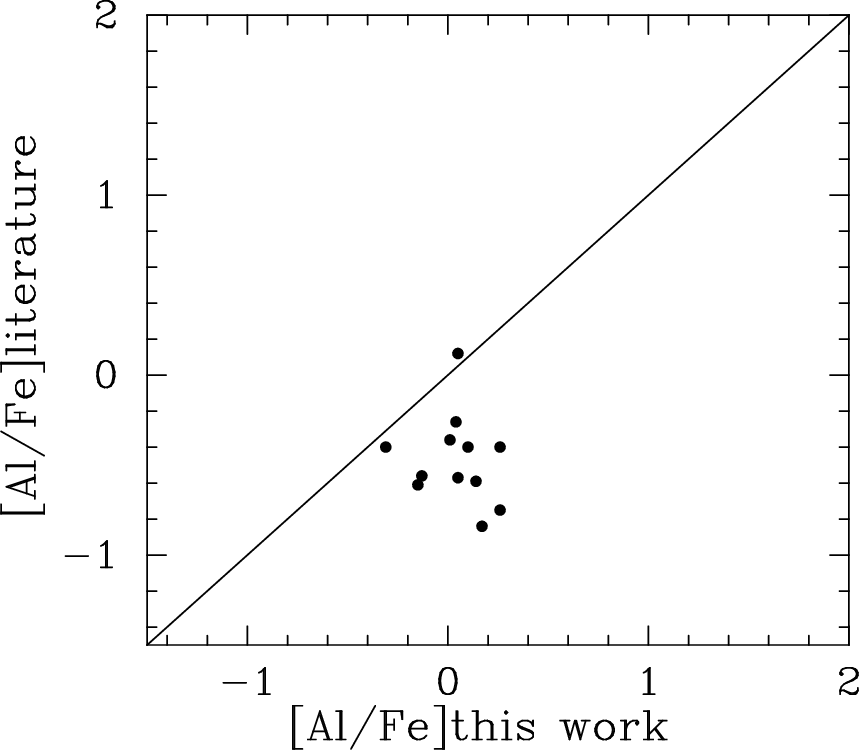}
   \includegraphics[width=6cm]{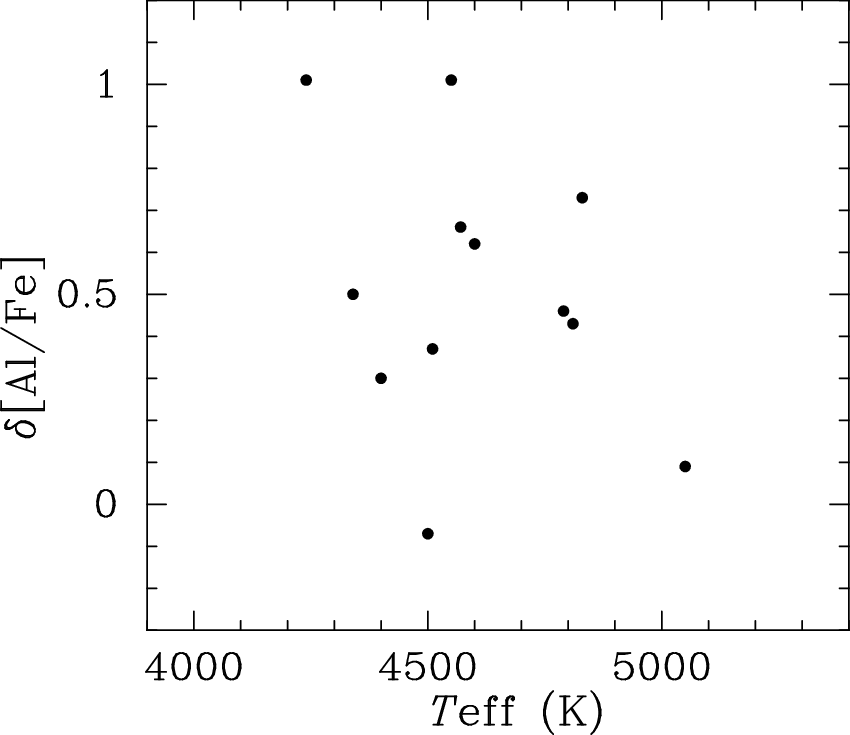}
   \includegraphics[width=6cm]{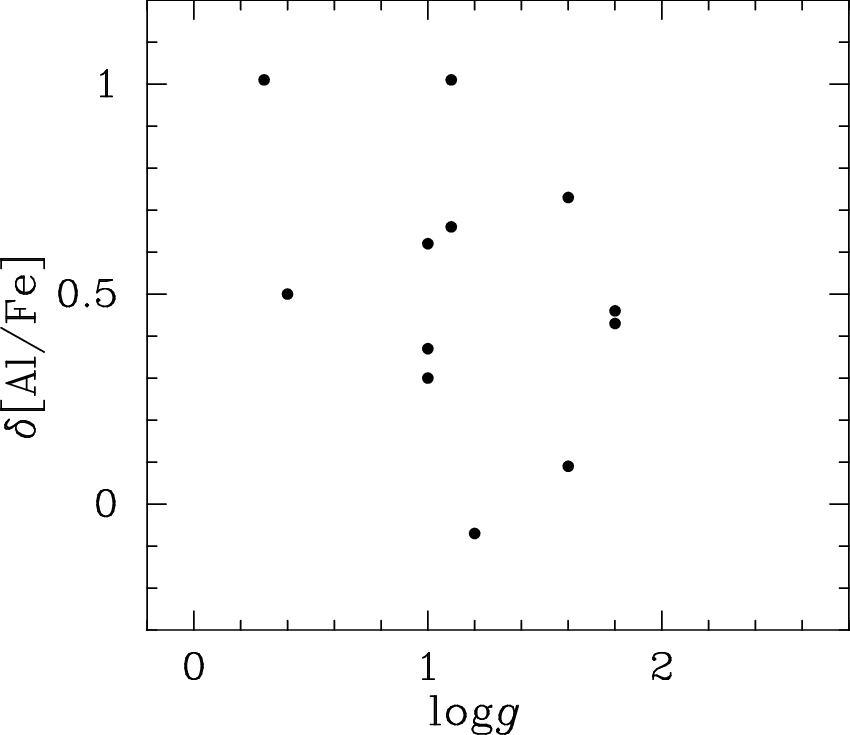}
   \includegraphics[width=6cm]{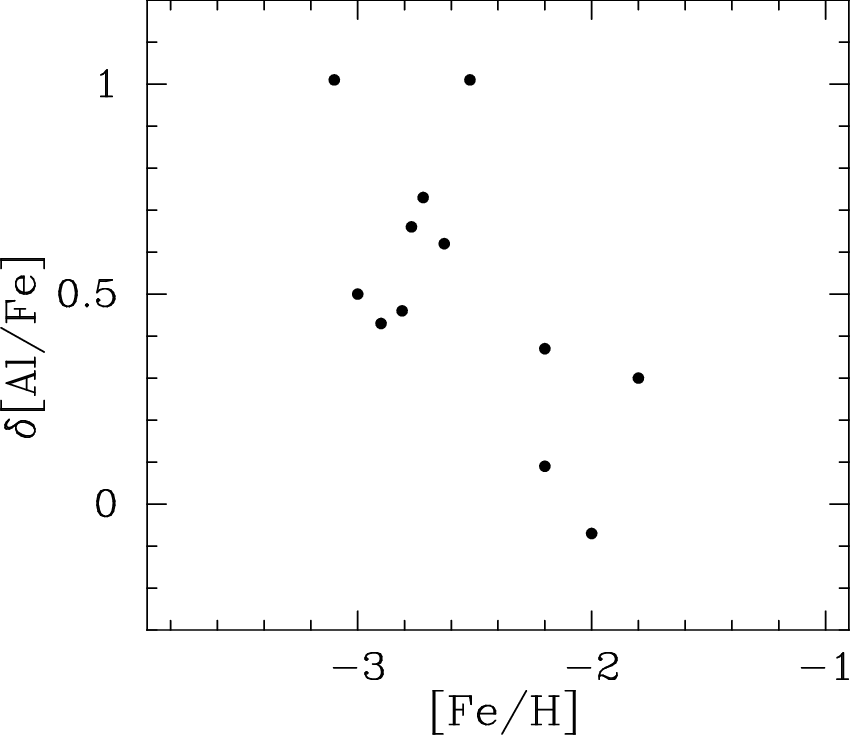}
 \end{center}
\caption{(top panel) Comparison of [Al/Fe] obtained by this work from the NIR spectra with those from optical lines in the literature. (2nd to 4th panels from the top) Differences of [Al/Fe] between this work and literature ($\delta$ [Al/Fe] = [Al/Fe]$_{\rm this work}-$[Al/H]$_{\rm literature}$), as functions of $T_{\rm eff}$, $\log g$, and [Fe/H]. {Alt text: A graph showing comparisons between abundance ratios [Al/Fe]. The x axis shows [Al/Fe] determined by the present work from $-1.5$ to $+2.0$ and the y axis shows those from the literature from $-1.5$ to $+2.0$. Other three graphs showing differences of [Al/Fe] values between this work and literature. The x axis shows stellar parameters and abundance ratios. The y axis shows the abundance differences.}}\label{fig:alcomp}
\end{figure}

Al abundances are determined for 27 stars from our NIR spectra. 
  The averages and standard deviations of [Al/Fe] are given in
  table~\ref{tab:ave}. The average values are clearly higher than
those obtained by previous studies for VMP stars from the optical Al
{\small I} 394.4~nm and/or 396.1~nm lines by on the order of
0.5--1.0~dex.  The Al abundance measurements from optical spectra have
been reported for a limited number of stars in our
sample. Figure~\ref{fig:alcomp} (top panel) compares the [Al/Fe] obtained by the
present work with the previous results from optical spectra for 12
stars (\cite{Honda2004ApJ}, \cite{Roederer2014AJ},
\cite{Holmbeck2018ApJ}, \cite{Johnson2002ApJS},
\cite{Fulbright2000AJ}, \cite{Barklem2005AA}, and
\cite{Li2013ApJ}). 


This discrepancy in [Al/Fe] could be due to NLTE effects, primarily
for the optical lines. \citet{Nordlander2017AA} have reported that the
NLTE corrections for the Al abundances determined from the 394.4~nm
and/or 396.1~nm lines are positive, and as large as 0.5~dex in VMP
giant stars ($T_{\rm eff}\leq 5000$ K, $\log g= 3$). On the other
hand, small negative NLTE corrections are predicted for the NIR line
at 1312.3~nm (\cite{Nordlander2017AA}). Larger effect is
  expected for the above two lines in the blue region for stars with lower metallicity. This might be found in the
  bottom panel of figure~\ref{fig:alcomp}, which shows a correlation
  between the abundance differences between this work and literature
  as a function of metallicity. It should be noted, however, that the Al
  abundances of the two stars with highest metallicity in the figure are determined
  from weaker Al I lines in 700-900~nm (HD8724:\cite{Fulbright2000AJ}; HD108577:\cite{Johnson2002ApJS}), for which the NLTE effects are
  expected to be much smaller than for the 394.4~nm and 396.1~nm
  lines. Further studies covering higher metallicity is required to
  examine the metallicity dependence of the NLTE effect. No clear correlation is found between the
  abundance differences and the stellar evolutionary status ($T_{\rm
    eff}$ and $\log g$) in figure~\ref{fig:alcomp}.

The discrepancy between the
Al abundances from NIR and optical lines is mostly resolved by the NLTE
corrections. Since the NLTE effect for the NIR lines is predicted to
be much smaller than those in the blue optical range, the results from the NIR
lines should better trace the Al abundance trend than the values from
the LTE analysis of optical lines. A comparison with 
chemical-evolution models is provided in \S~\ref{sec:disc}.

\subsection{Sr}

\begin{figure}
 \begin{center}
   \includegraphics[width=8cm]{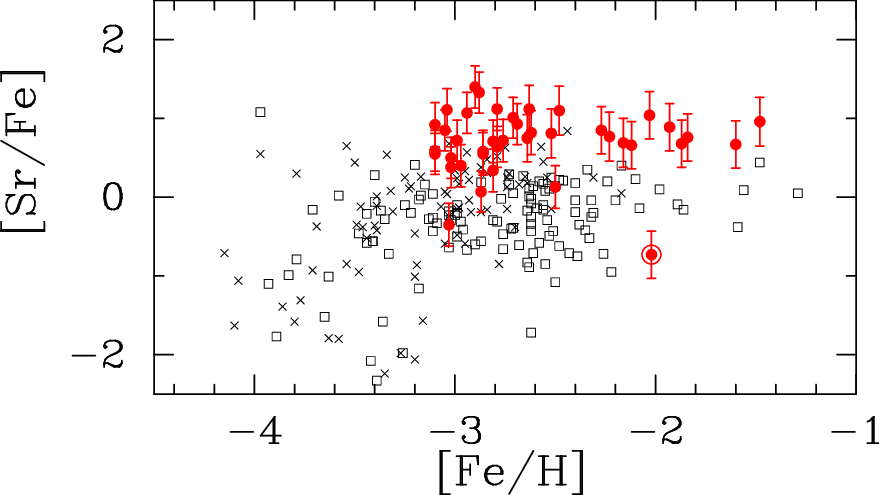}
 \end{center}
 \caption{The same as figure~\ref{fig:mgsi}, but for [Sr/Fe]. The
   abundance trend is not shown because the Sr abundances determined
   from NIR lines are severely affected by NLTE effects (see
   text).{Alt text: A graph showing abundance ratios [Sr/Fe]. The x axis shows [Fe/H] from $-4.5$ to $-1.0$ and the y axis shows the abundance ratios from $-2.5$ to +2.5.}}\label{fig:sr}
\end{figure}

\begin{figure}
 \begin{center}
   \includegraphics[width=6cm]{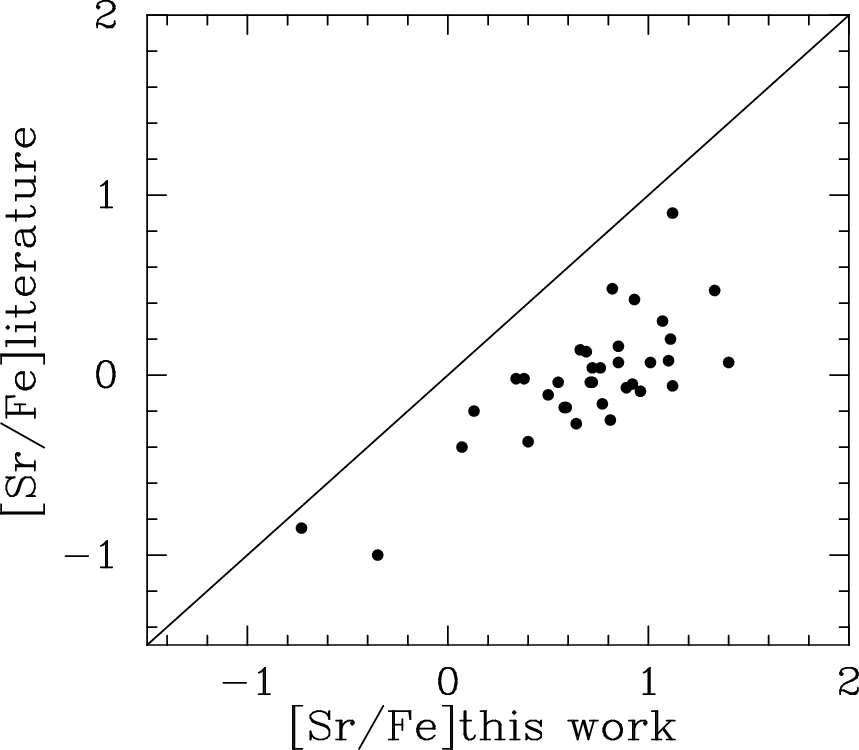}
   \includegraphics[width=6cm]{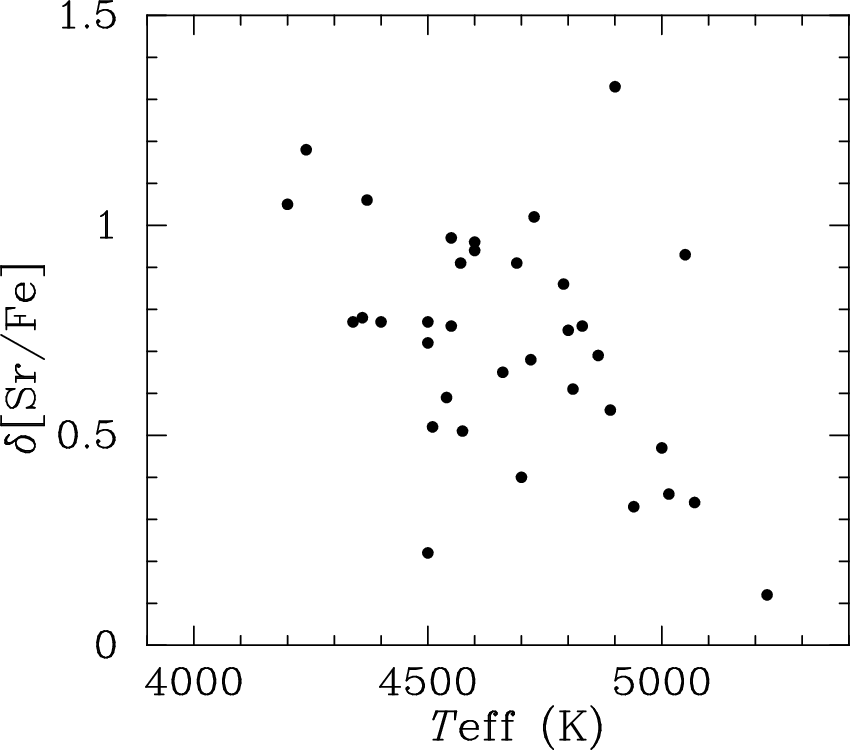}
   \includegraphics[width=6cm]{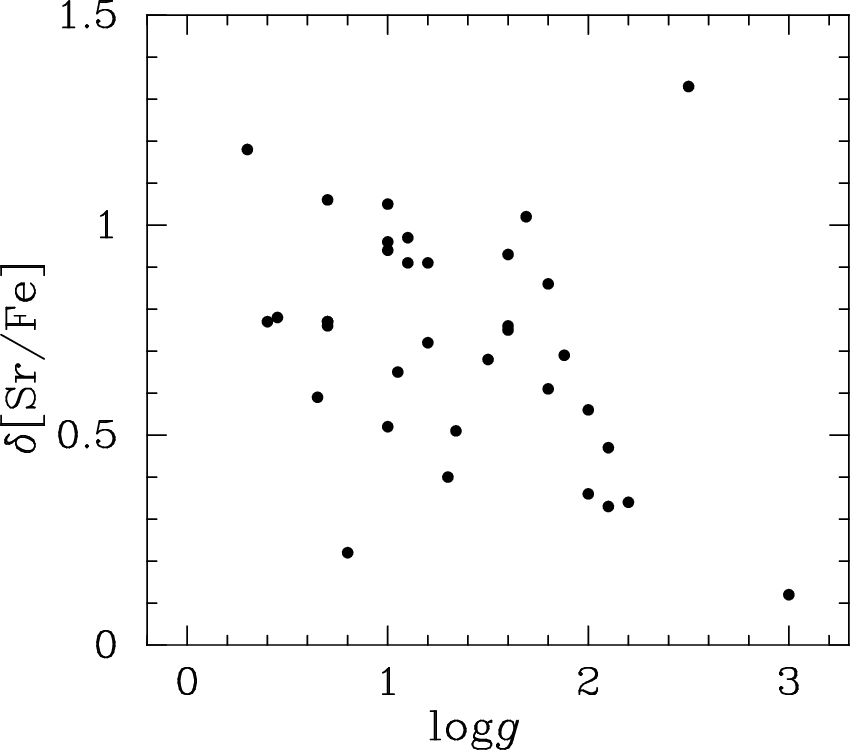}
   \includegraphics[width=6cm]{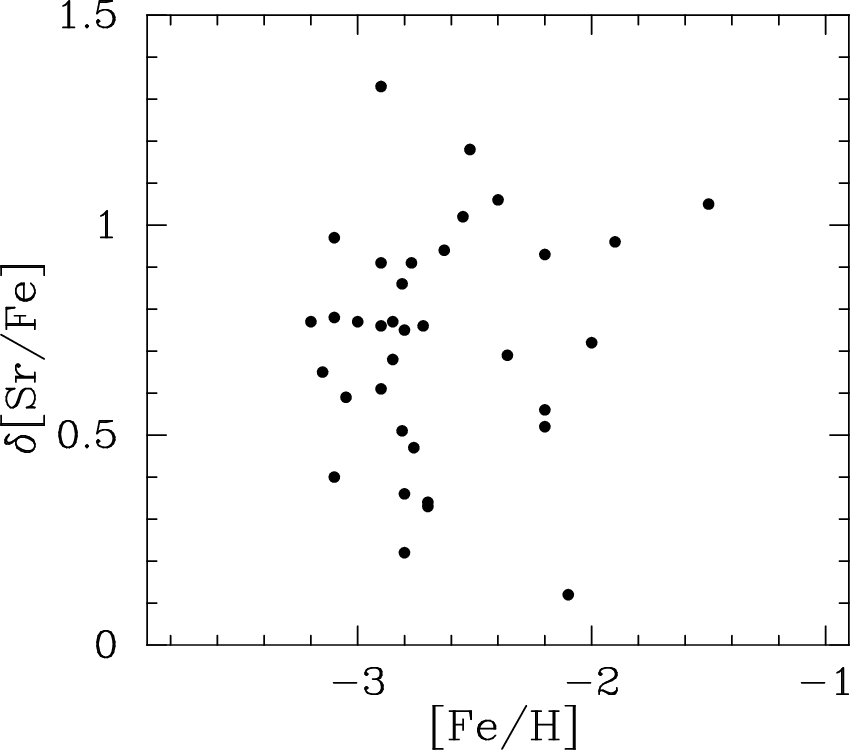}
 \end{center}
 \caption{The same as figure~\ref{fig:srcomp}, but for Sr.}\label{fig:srcomp}
\end{figure}

The Sr abundances are determined from the Sr {\small II} triplet in
the $J$-band (1003.7~nm, 1032.7~nm, and 1091.5~nm) in the present work. Figure~\ref{fig:sr} shows [Sr/Fe], as a
function of [Fe/H], for our sample and those determined by previous
studies from the optical resonance lines. The [Sr/Fe] values obtained
by the present work are systematically higher than those from
literature, as found in table~\ref{tab:ave}. Figure~\ref{fig:srcomp} (top panel) compares [Sr/Fe] values from this
work and from the literature, indicating that the discrepancy is
systematic, and is not clearly dependent on the [Sr/Fe] values.

\citet{Aoki2022PASJ} have already reported the discrepancy of the Sr
abundances from NIR and optical lines for HD~4306 and HD~221170. They
demonstrated that the Sr abundance of the Sun derived by their analysis
of the Sr {\small II} triplet reproduces the value obtained by
\citet{Grevesse2015AA}, concluding that the line data are not the
reason for the discrepancy. On the other hand, large NLTE effects on
the Sr II line formation have been reported for cool red giants (e.g.,
\cite{Andrievsky2011AA, Bergemann2012AA}). Hence,
below we inspect the correlations between this abundance discrepancy and
the stellar parameters in order to examine the NLTE models. 

Figure~\ref{fig:srcomp} (lower three panels) shows the discrepancy of the [Sr/Fe] abundance
ratios between those from the NIR and optical lines, as a function of stellar
parameters ($T_{\rm eff}$, $\log g$, metallicity). There are clear dependencies of the discrepancy on 
$T_{\rm eff}$ and $\log g$: the discrepancy is larger in cooler and more
evolved red giants. The discrepancy is as large as 1~dex, which is found for giants with
the lowest temperature and gravity, for which a larger NLTE effect is
predicted. An exception is 2MASS J0643+5934, which exhibits a discrepancy of
1.3~dex, although the temperature and gravity  (4900~K
and 2.5, respectively) are not very low in our sample. This star was studied by \citet{Bandyopadhyay2022ApJ} for the optical lines. Their Table 2 reports that the $T_{\rm eff}$'s determined by the indicators other than that from the spectroscopic analysis of Fe I lines are lower than the adopted value (4900~K), e.g. 4600~K from the color $V-K$. Hence, the $T_{\rm eff}$ of this star might be lower than that adopted in their work. On the other hand, the $\log g$ values derived from all the methods, including that from Gaia parallax, are higher than 2.2 (their Table 3), suggesting that the $\log g$ value is robust. Hence, no clear reason for the large discrepancy of the Sr abundance found for 2MASS J0643+5934 is identified.

\citet{Andrievsky2011AA} present large negative NLTE corrections for [Sr/Fe] abundance ratios
from  NIR infrared lines and smaller positive corrections for those from the
optical resonance lines for cool red giants ($T_{\rm eff}=4500$~K and
5000~K). The difference of the corrections between the NIR and
optical lines are as large as 0.5~dex.
Similar NLTE corrections are demonstrated by \citet{Nordlander2017AA} for metal-poor giants. The discrepancy found in the observational results from NIR lines (this work) and optical
ones (the literature) are at least qualitatively explained by NLTE
effects, supporting the NLTE calculations by the above work. For better understanding of the formation of optical and NIR Sr lines, further studies for individual Sr lines in stars with a variety of stellar parameters are desired. 

\section{Discussion and concluding remarks}\label{sec:disc}

We have determined Mg and Si abundances for 44 metal-poor stars from
measurements of spectral lines identified in high-resolution
NIR spectra.  We found no statistically significant scatter for both [Mg/Fe] and
[Si/Fe], excluding one or two outliers. Predictions of the yields from
simulations of massive star evolution and core-collapse supernovae (e.g., \cite{Tominaga2007ApJ}; \cite{Heger2010ApJ})
indicate over-abundances of Mg and Si with respect to Fe in general,
with dependence on the initial stellar masses.

The overall trend of the abundance ratios of Mg and Si are well-reproduced by the chemical-evolution models of
\citet{Kobayashi2020ApJ}. The averages of [Mg/Fe] for stars in the
ranges $-3.5 < $ [Fe/H] $ < -2.5$ and $-2.5 < $ [Fe/H] $ < -1.5$ of our sample are +0.46 and
+0.43, respectively, whereas the [Mg/Fe] values of their chemical-evolution models at [Fe/H] $ = -3.0$ and $-2.0$ are both +0.45. We note that the $\alpha$-poor star BD+80$^{\circ}$245 is excluded from the
statistics. We also note that [Mg/Fe] exhibits a decreasing trend with increasing metallicity,
while an almost constant over-abundance
of Mg is predicted by chemical-evolution models. The slope reported
in the previous section ($\delta$[Mg/Fe]/$\delta$[Fe/H]$ = -0.10\pm$0.05) is estimated including the two most metal-poor
stars with [Fe/H] $ < -3.5$ (BD+44$^{\circ}$493 and HE~1116-0634) that have relatively high [Mg/Fe]. If these two stars are excluded from the statistics, the slope is not significant: $\delta$[Mg/Fe]/$\delta$[Fe/H]$=-0.05\pm 0.05$. 


The averages of [Si/Fe] in the ranges $-3.5 < $ [Fe/H] $ < -2.5$ and
$-2.5 < $ [Fe/H] $ < -1.5$ of our sample are +0.57 and +0.50, respectively, whereas the
[Si/Fe] values of the chemical-evolution models of \citet{Kobayashi2020ApJ} at [Fe/H] $ = -3.0$ and
$-2.0$ are +0.58 and +0.51, respectively, exhibiting almost perfect
agreement. The decreasing trend of [Si/Fe] with increasing metallicity ($\delta$[Si/Fe]/$\delta$[Fe/H]$ = -0.08\pm$0.04)
is also reproduced by the model. BD+80$^\circ$245 is excluded from these
statistics. We note for completeness that the trend does not change if the three most metal-poor stars are excluded.


A remarkable result of our study is the small scatter of [Si/Fe]. The
small scatter of [Mg/Fe] found by previous studies is also
confirmed. Such small scatter in very metal-poor stars is not
necessarily expected from the nucleosynthesis in early generations of
massive stars and supernovae, because production environments of the
three elements are different: Mg is produced in C burning during
stellar evolution; Si is mostly produced in explosive O burning; Fe is
produced by the complete Si burning. The ejected mass of Si and Fe depends on
the fallback (e.g., \cite{Woosley1995ApJS, Francois2004AA}). From the small scatter of [Mg/Fe] found from their
high-quality spectra, \citet{Cayrel2004AA} discuss that mixing of
interstellar matter is efficient even at low metallicity (see also
\cite{Francois2004AA}). On the other hand, very large scatter of
abundance ratios of neutron-capture elements, e.g., [Sr/Fe] and
[Ba/Fe], are found in very metal-poor stars. This indicates relatively
small variations of [Mg/Fe] and [Si/Fe] provided by individual
supernovae (e.g., \cite{Scannapieco2022MNRAS}). Taking account of the
dependence of those abundance ratios on progenitor mass, the mass
ranges of progenitors that made dominant contributions to very
metal-poor stars might be limited. As larger scatter is expected for
those abundance ratios in most metal-poor stars (e.g., [Fe/H]$<-3.5$)
because they would more clearly reflect the yields of individual
supernovae, extension of measurements from NIR spectra to lower
metallicity will be useful to examine this possibility.


There are two outliers: BD+80$^{\circ}$245 exhibits very low [Mg/Fe]
and [Si/Fe], and LAMOST~J2217+2104 has a significantly high
[Mg/Fe]. The abundance anomalies in these stars have been identified
by previous studies based on optical spectra. Our study of NIR spectra
confirms the anomaly of the Si abundance ratio of BD+80$^{\circ}$245 with higher precision than the previous studies (e.g., \cite{Ivans2003ApJ}). The low abundance ratios of both Mg and Si, as well as of Ca and Ti that were studied by previous work, suggest that Fe is enhanced in this object most likely by larger contributions of type Ia supernovae than other stars with similar metallicity.

The large excess of Mg, along with the moderate excess of Si in LAMOST~J2217+2104, suggests that the mechanism that enhances Mg
and Si in the progenitor of this object, which could be a supernova
explosion of a first-generation massive star, is more or less
different from that of core-collapse supernovae that have produced the
typical abundance ratios found in metal-poor stars. This is a CEMP star with large excesses of N and O. \citet{Aoki2018PASJ} discuss that the elemental abundance pattern from C is best explained by the explosion of a massive star with 25~M$_{\odot}$, which is not particularly different from typical mass of progenitors expected for bulk of metal-poor stars. Hence, there could be other unique features, e.g., rapid rotation or binarity, in the progenitor of this star. 
We note that our sample includes two CEMP stars with no excess of s-process elements, CS~29502--092 and CS~31014--067 \citep{Aoki2002ApJ}. Their abundance ratios of [Mg/Fe] and [Si/Fe], as well as [Sr/Fe], are indistinguishable from those of other carbon-normal stars.

Our sample includes two r-II stars: CS~31082-001, and HE~1523-0901. Although the Mg and Si abundance ratios of HE~1523-0901 ([Mg/Fe] = +0.25 and [Si/Fe] = +0.39) are slightly lower than the averages, the departure is not statistically significant. The abundance ratios of CS~31082-001 agrees well with the averages. Our result supports the suggestions from previous studies based on optical spectra that there is no unique feature in the $\alpha$/Fe abundance ratios in $r$-process enhanced stars (e.g., \cite{Roederer2009AJ}). We note that the Sr abundances of these two stars are enhanced.  

The [Al/Fe] values in our work are flat within the errors
($\delta$[Al/Fe]/$\delta$[Fe/H]$ = -0.02\pm$0.09). An increasing trend
of [Al/Fe] with increasing metallicity is expected in [Fe/H]$<-1$
because the production of odd-Z elements depends on the surplus of
neutrons from $^{22}$Ne \citep{Kobayashi2020ApJ}. Although such a
trend is not found in our result, it is not excluded taking account of
the errors due to small number of available lines and possible NLTE
effects that are still uncertain. For the same reason, the size of the
scatter is not well constrained by the present work. More
comprehensive NLTE analyses of both optical and NIR Al lines will be
useful to constrain the chemical evolution models.

This work demonstrates that high-resolution NIR spectra are useful to determine abundances of key elements in metal-poor stars. The Mg and Si abundances are well-determined by measurements of a larger number of spectral lines than available in the optical range. Measurements of NIR lines also contribute to examining the NLTE effects for Sr by providing additional transitions to the previous studies based on optical spectra. Our study focused on bright metal-poor stars that cover mostly the metallicity range with [Fe/H]$>-3.5$. The observation with IRD is, however, able to obtain sufficiently good spectra of stars with lower metallicty. For instance, Mg and Si abundances are well-determined for LAMOST~J2217--2104 with [Fe/H]$=-3.9$ and $V=13.4$. Further NIR observations of such most metal-poor stars will contribute to constraining the abundance trends and scatter produced by the first generations of stars.  

\begin{ack}

This research is based on data collected at Subaru Telescope, which is operated by the National Astronomical Observatory of Japan. We are honored and grateful for the opportunity of observing the Universe from Maunakea, which has the cultural, historical and natural significance in Hawaii.
We are grateful to Drs. Nozomu Tominaga and Miho N. Ishigaki for their useful comments on stellar yields and chemical evolution. 
This work was supported in part by 
the National Science Foundation under Grant No. OISE-1927130 (IReNA). W.A. is supported by JSPS KAKENHI grant No. 21H04499.
T.C.B. acknowledges partial support from grant PHY 14-30152
(Physics Frontier Center/JINA-CEE), and OISE-1927130: The International Research Network for Nuclear Astrophysics (IReNA), awarded by the U.S. National Science Foundation. M.T. is supported by JSPS KAKENHI grant No.24H00242.
The work of V.M.P. is supported by NOIRLab, which is managed by the Association of Universities for Research in Astronomy (AURA) under a cooperative agreement with the National Science Foundation.

\end{ack}

\clearpage

\clearpage
   

\clearpage
\begin{longtable}{lllrrr}
  \caption{Objects and Observations}\label{tab:obj}
\hline\noalign{\vskip3pt} 
Star & Gaia ID & Obs. Date & Exp. Time & S/N  & S/N \\
     &         &  (UT)     &  (sec) &  (1050nm) & (1600nm) \\
\hline\noalign{\vskip3pt} 
\endfirsthead      
\endhead
\hline\noalign{\vskip3pt} 
\endfoot
HD~4306 & 2473977823368495360 & Jul. 25, 2020 & 180 & 57 & 78 \\
HD~6268 & 5033960575837265920 & Nov. 8, 2022 & 120 & 75 & 115 \\
HD~8724 & 2593188145361730432 & Nov. 8, 2022 & 120 & 78 & 116 \\
HD~13979 & 5118012879660872832 & Nov. 8, 2022 & 300 & 91 & 132 \\
HD~85773         & 5660878702718068864 & Mar. 15, 2022 & 300 & 120 & 182 \\
HD~88609         & 851658896305185024 & Mar. 15, 2022 & 180 & 110 & 165 \\
HD~107752         & 3907404496774653312 & Mar. 15, 2022 & 630 & 109 & 154 \\
HD~108577         & 3907872613852996864 & Mar. 15, 2022 & 420 & 93 & 130 \\
HD~110184         & 3710786979933623424 & Mar. 15, 2022 & 120 & 98 & 155 \\
HD~115444         & 1474455748663044736 & Mar. 15, 2022 & 300 & 116 & 175 \\
HD~118055     & 3603949359207346048 & Mar. 15, 2022 & 300 & 164 & 257 \\
HD~122563         & 3723554268436602240 & Mar. 15, 2022 & 60 & 202 & 297 \\
HD~187111         & 4189688954569244032 & Nov. 8, 2022 & 60 & 115 & 186 \\
HD~204543         & 2672702897842335872 & Nov. 8, 2022 & 180 & 144 & 214 \\
HD~221170 & 2869759781250083200 & Jul. 25, 2020 & 120 & 134 & 230 \\
HD~237846         & 1049376272667191552 & Mar. 15, 2022 & 600 & 103 & 141 \\
BD$+03^{\circ}2782$          & 3712682602403001728 & Mar. 15, 2022 & 600 & 97 & 165 \\
BD$+30^{\circ}2611$          & 1275876252107941888 & Mar. 15, 2022 & 180 & 110 & 173 \\
BD$+44^{\circ}493$ & 341511064663637376 & Jul. 25, 2020 & 900 & 187 & 253 \\
BD$+80^{\circ}245$          & 1139085117140997120 & Mar. 15, 2022 & 900 & 69 & 89 \\
BD$-02^{\circ}5957$ & 2643630302870307328 & Nov. 8, 2022 & 900 & 58 & 86 \\
BD$-07^{\circ}2674$       & 5755996908175855232 & Mar. 15, 2022 & 600 & 100 & 149 \\
BD$-14^{\circ}5890 $       & 6888560851915102464 & Nov. 8, 2022 & 600 & 88 & 124 \\
BD$-15^{\circ}5781$          & 6887351938881045504 & Nov. 8, 2022 & 900 & 48 & 67 \\
BD$-18^{\circ}271$          & 2450584991932903168 & Nov. 8, 2022 & 300 & 92 & 148 \\
BD$-18^{\circ}5550$          & 6867802519062194560 & Nov. 8, 2022 & 540 & 116 & 188 \\
BD$-20^{\circ}6008$          & 6857966322398347008 & Nov. 8, 2022 & 360 & 79 & 115 \\
CS~29502--092          & 2629500925618285952 & Nov. 8, 2022 & 1800 & 61 & 91 \\
CS~30314--067         & 6779790049231492096 & Nov. 8, 2022 & 1800 & 87 & 130 \\
CS~31082--001          & 2451773937663757696 & Nov. 8, 2022 & 2400 & 58 & 120 \\
HE~1116--0634         & 3783989199935482624 & Mar. 15, 2022 & 1800 & 91 & 137 \\
HE~1320--1339         & 3608447014600045824 & Mar. 15, 2022 & 900 & 91 & 132 \\
HE~1523--0901         & 6317828550897175936 & Mar. 15, 2022 & 480 & 54 & 101 \\
LAMOST J003251.47+410749.0 & 381063654713703040 & Nov. 8, 2022 & 900 & 22 & 35 \\
LAMOST J004012.52+272924.7 & 2809551447929127040 & Nov. 8, 2022 & 1350 & 84 & 121 \\
LAMOST J074847.30+461308.4 &  927340854329737216 & Mar. 15, 2022 & 900 & 60 & 89 \\
(TYC  3407-1352-1) & & &&& \\
LAMOST J210958.01+172544.0 & 1788003032879354752 & Nov. 8, 2022 & 600 & 69 & 103 \\
LAMOST J211401.52$-$061610.3 & 6910940758263238912  & Nov. 8, 2022 & 900 & 81 & 117 \\
LAMOST J221750.59+210437.2 & 1778804140643594240 & Jul. 25, 2020 & 5400 & 73 & 110 \\
LAMOST J234759.59+285130.7 & 2866981452805072640 & Nov. 8, 2022 & 1350 & 71 & 94 \\
2MASS J06430186+5934309 & 1001955470034395776 & Mar. 15, 2022 & 900 & 22 & 30 \\
(TYC  3778-529-1 ) & & &&& \\
2MASS J09544277+5246414 & 828438619475671936  & Mar. 15, 2022 & 420 & 98 & 150 \\
(TYC  3814-1598-1) & & &&& \\
2MASS J20165357$-$0503592 & 4219051451239614080 & Nov. 8, 2022 & 1800 & 52 & 85 \\
(UCAC4  425-121652) & & &&& \\
2MASS J21454816+1249124 & 1767214390437753600 & Nov. 8, 2022 & 1800 & 63 & 96 \\
(UCAC4  515-137892) & & &&& \\
\end{longtable}

\begin{table}
  \tbl{Line Data and Equivalent Widths}{%
    \begin{tabular}{llrrrr}
      \hline
      Star & Species & Wavelength (nm) & $\log gf$ & $\chi$(eV) & $W$(pm) \\
\hline
HD~4306          &  Sr II & 1003.665 &$-1.310$ &1.805 & 1.33 \\
HD~4306          &  Sr II & 1032.731 &$-0.350$ &1.839 & 6.96\\
HD~4306          &  Si I  & 1037.126 &$-0.705$ &4.930 & 3.06\\
HD~4306          &  Si I  & 1058.514 &$ 0.012$ &4.954 & 7.73\\
HD~4306          &  Si I  & 1060.343 &$-0.305$ &4.930 & 5.19\\
     \hline
    \end{tabular}}\label{tab:ew}
  \begin{tabnote}
    The whole data are available at PASJ online.
  \end{tabnote}  
\end{table}

\footnotesize
\begin{longtable}{lrrrrrrrrrrrr}
  \caption{Stellar Parameters and Fe Abundances}\label{tab:param}
\hline\noalign{\vskip3pt} 
Object & $T_{\rm eff}$  & $\log g$ &  [Fe/H] &  $v_{\rm turb}$ & ref. \footnotemark[$*$] & [Fe/H]$_{\rm ref}$ & [Fe/H]$_{\rm opt}$ & $N_{\rm opt}$ & $\sigma_{\rm opt}$ & [Fe/H]$_{\rm IRD}$ & $N_{\rm IRD}$& [Fe/H]$_{\rm fin}$ \\
 & (K)  &  &   &  (km~s$^{-1}$) &   &  &  & &   &  & &  \\
\hline\noalign{\vskip3pt} 
\endfirsthead      
\endhead
\hline\noalign{\vskip3pt} 
\endfoot
\hline\noalign{\vskip3pt} 
\multicolumn{2}{@{}l@{}}{\hbox to0pt{\parbox{160mm}{\footnotesize
\hangindent6pt\noindent
\hbox to6pt{\footnotemark[$*$]\hss}\unskip%
  References -- 1:\citet{Honda2004ApJ}; 2:\citet{Burris2000ApJ}; 3:\citet{Roederer2014AJ}; 4:\citet{Ishigaki2012ApJ}; 5:\citet{Ishigaki2013ApJ}; 6. \citet{Honda2007ApJ}; 7:\citet{Fulbright2000AJ}; 8:\citet{Honda2006ApJ}; 9:\citet{Aoki2008PASJ}; 10:\citet{Ivans2006ApJ}; 11:\citet{Ito2013ApJ}; 12: \citet{Ivans2003ApJ}; 13:\citet{Hansen2018ApJ}; 14:\citet{Cain2018ApJ}; 15:\citet{Aoki2002ApJ}; 16:\citet{Hollek2011ApJ}; 17:\citet{Li2022ApJ}; 18:\citet{Aoki2018PASJ}; 19:\citet{Bandyopadhyay2022ApJ}; 20:\citet{Holmbeck2018ApJ}; 21:\citet{Placco2019ApJ}
  \hss}}}
\endlastfoot 
HD~4306 & 4810 & 1.8 & $-$2.9 & 1.6 & 1 & $-$2.89 & $-$3.02 & 57 & 0.09 & $-$2.78 & 1 & $-$3.02 \\
HD~6268         & 4600 & 1.0 & $-$2.6 & 2.1 & 1 & $-$2.62 & $-$2.71 & 49 & 0.10 & $-$2.36 & 5 & $-$2.71 \\
HD~8724 & 4500 & 1.2 & $-$2.0 & 1.8 & 2 & $-$1.84 &  &  &  & $-$1.92 & 22 & $-$1.84 \\
HD~13979 & 4830 & 1.6 & $-$2.7 & 1.6 & 3 & $-$2.97 & $-$2.86 & 94 & 0.08 & $-$2.73 & 1 & $-$2.86 \\
HD~85773         & 4370 & 0.7 & $-$2.4 & 2.0 & 4,5 & $-$2.43 & $-$2.52 & 57 & 0.09 & $-$2.42 & 11 & $-$2.52 \\
HD~88609         & 4550 & 1.1 & $-$3.1 & 2.4 & 1,6 & $-$3.06 & $-$3.10 & 51 & 0.16 & $-$2.94 & 2 & $-$3.10 \\
HD~107752         & 4800 & 1.6 & $-$2.8 & 1.9 & 4,5 & $-$2.77 & $-$2.81 & 80 & 0.13 & $-$2.75 & 2 & $-$2.81 \\
HD~108577         & 5050 & 1.6 & $-$2.2 & 1.9 & 4,5 & $-$2.23 & $-$2.23 & 76 & 0.09 & $-$2.13 & 6 & $-$2.23 \\
HD~110184         & 4240 & 0.3 & $-$2.5 & 2.1 & 1 & $-$2.51 & $-$2.63 & 25 & 0.11 & $-$2.55 & 11 & $-$2.63 \\
HD~115444         & 4720 & 1.5 & $-$2.9 & 1.7 & 1 & $-$2.84 & $-$2.99 & 56 & 0.12 & $-$2.94 & 4 & $-$2.99 \\
HD~118055     & 4400 & 1.0 & $-$1.8 & 2.3 & 7 & $-$1.80 & $-$1.87 & 22 & 0.06 & $-$1.85 & 29 & $-$1.87 \\
HD~122563         & 4570 & 1.1 & $-$2.8 & 2.1 & 1,8 & $-$2.76 & $-$2.79 & 63 & 0.14 & $-$2.63 & 5 & $-$2.79 \\
HD~187111         & 4450 & 1.1 & $-$1.6 & 1.9 & 7 & $-$1.60 & $-$1.60 & 18 & 0.08 & $-$1.60 & 31 & $-$1.60 \\
HD~204543         & 4600 & 1.0 & $-$1.9 & 2.2 & 9 & $-$1.86 & $-$1.93 & 52 & 0.14 & $-$1.91 & 17 & $-$1.93 \\
HD~221170 & 4510 & 1.0 & $-$2.2 & 2.1 & 10 & $-$2.19 & $-$2.12 & 101 & 0.13 & $-$2.22 & 13 & $-$2.12 \\
HD~237846         & 5015 & 2.0 & $-$2.8 & 1.6 & 4,5 & $-$2.81 & $-$2.81 & 80 & 0.15 & $-$2.76 & 3 & $-$2.81 \\
BD$+03^{\circ}2782$          & 4600 & 1.5 & $-$2.0 & 1.8 & 2 & $-$2.03 &  &  &  & $-$2.04 & 14 & $-$2.03 \\
BD$+30^{\circ}2611$          & 4200 & 1.0 & $-$1.5 & 1.8 & 4,5 & $-$1.53 & $-$1.48 & 20 & 0.08 & $-$1.47 & 38 & $-$1.48 \\
BD$+44^{\circ}493$ & 5430 & 3.4 & $-$3.8 & 1.3 & 11 & $-$3.80 & $-$3.83 & 179 & 0.13 &   &  & $-$3.83 \\
BD$+80^{\circ}245$          & 5225 & 3.0 & $-$2.1 & 1.3 & 12 & $-$2.07 & $-$2.02 & 198 & 0.11 & $-$2.00 & 6 & $-$2.02 \\
BD$-02^{\circ}5957$ & 4360 & 0.5 & $-$3.1 & 2.4 & 13 & $-$3.13 & $-$3.05 & 79 & 0.16 & $-$3.09 & 2 & $-$3.05 \\
BD$-07^{\circ}2674$       & 4500 & 0.7 & $-$3.2 & 2.3 & 14 & $-$3.16 & $-$3.10 & 255 & 0.15 & $-$2.94 & 5 & $-$3.10 \\
BD$-14^{\circ}5890$        & 4890 & 2.0 & $-$2.2 & 1.7 & 4,5 & $-$2.16 & $-$2.16 & 83 & 0.10 & $-$2.12 & 6 & $-$2.16 \\
BD$-15^{\circ}5781$          & 4550 & 0.7 & $-$2.9 & 1.7 & 3 & $-$2.87 & $-$2.76 & 102 & 0.08 & $-$2.72 & 3 & $-$2.76 \\
BD$-18^{\circ}271$          & 4230 & 0.4 & $-$2.6 & 2.5 & 4,5 & $-$2.58 & $-$2.64 & 50 & 0.09 & $-$2.55 & 15 & $-$2.64 \\
BD$-18^{\circ}5550$         & 4660 & 1.1 & $-$3.2 & 1.6 & 3 & $-$3.20 & $-$3.03 & 107 & 0.09 & $-$2.69 & 2 & $-$3.03 \\
BD$-20^{\circ}6008$          & 4540 & 0.7 & $-$3.1 & 1.7 & 3 & $-$3.00 & $-$2.86 & 86 & 0.08 & $-$2.72 & 1 & $-$2.86 \\
CS~29502--092          & 5000 & 2.1 & $-$2.8 & 1.8 & 15 & $-$2.76 & $-$2.87 & 171 & 0.17 & $-$2.93 & 3 & $-$2.87 \\
CS~30314--067         & 4400 & 0.7 & $-$2.9 & 2.5 & 15 & $-$2.85 & $-$2.97 & 111 & 0.15 & $-$2.95 & 3 & $-$2.97 \\
CS~31082--001          & 4790 & 1.8 & $-$2.8 & 1.9 & 1 & $-$2.80 & $-$2.88 & 48 & 0.09 & $-$2.71 & 4 & $-$2.88 \\
HE~1116--0634         & 4400 & 0.1 & $-$3.7 & 2.4 & 16 & $-$3.73 & $-$3.66 & 86 & 0.18 & $-$3.68 & 2 & $-$3.66 \\
HE~1320--1339         & 4690 & 1.2 & $-$2.9 & 1.8 & 3 & $-$2.93 & $-$3.04 & 107 & 0.07 & $-$3.12 & 2 & $-$3.04 \\
HE~1523--0901         & 4500 & 0.8 & $-$2.8 & 2.4 & 13 & $-$2.83 & $-$2.79 & 74 & 0.14 & $-$2.87 & 3 & $-$2.79 \\
LAMOST J0032+4107 & 5070 & 2.2 & $-$2.7 & 1.8 & 17 & $-$2.69 & $-$2.62 & 123 & 0.11 & $-$2.74 & 1 & $-$2.62 \\
LAMOST J0040+2729 & 4574 & 1.3 & $-$2.8 & 2.2 & 17 & $-$2.80 & $-$2.69 & 130 & 0.11 & $-$2.70 & 2 & $-$2.69 \\
LAMOST J0748+4613 & 4700 & 1.3 & $-$3.1 & 2.2 & 17 & $-$3.10 & $-$3.02 & 131 & 0.12 & $-$2.99 & 2 & $-$3.02 \\
LAMOST J2109+1725 & 4940 & 2.1 & $-$2.7 & 1.7 & 17 & $-$2.57 & $-$2.50 & 146 & 0.10 & $-$2.39 & 1 & $-$2.50 \\
LAMOST J2114$-$0616 & 4727 & 1.7 & $-$2.6 & 1.8 & 17 & $-$2.55 & $-$2.48 & 105 & 0.09 & $-$2.25 & 3 & $-$2.48 \\
LAMOST J2217+2104 & 4500 & 0.9 & $-$3.9 & 2.3 & 18 & $-$3.90 & $-$3.93 & 53 & 0.16 &  &  & $-$3.93 \\
LAMOST J2347+2851 & 4864 & 1.9 & $-$2.4 & 1.8 & 17 & $-$2.36 & $-$2.27 & 139 & 0.10 & $-$2.22 & 2 & $-$2.27 \\
2MASS J0643+5934 & 4900 & 2.5 & $-$2.9 & 1.5 & 19 & $-$2.90 &  &  &  &  &  & $-$2.90 \\
2MASS J0954+5246       & 4340 & 0.4 & $-$3.0 & 2.3 & 20 & $-$2.99 & $-$2.94 & 104 & 0.11 & $-$2.78 & 6 & $-$2.94 \\
2MASS J2016$-$0507 & 4585 & 0.9 & $-$2.9 & 2.9 & 13 & $-$2.89 & $-$2.81 & 80 & 0.18 & $-$2.72 & 5 & $-$2.81 \\
2MASS J2145+1249 & 4580 & 0.5 & $-$3.3 & 2.7 & 21 & $-$3.10 &  &  &  & $-$3.09 & 3 & $-$3.10 \\
\end{longtable}

\begin{table}
  \tbl{Abundance Sensitivity to Stellar Parameters}{%
    \begin{tabular}{lcccccc}
      \hline
Element & $\Delta T_{\rm eff}$ & $\Delta \log g$ & $\Delta v_{\rm turb}$ & $\Delta$[Fe/H] & $\sigma_{\log\epsilon}$ & $\sigma_{\rm [X/Fe]}$ \\
        & (+100K) & (+0.3~dex) & (+0.3~km~s$^{-1}$) & (+0.3~dex) &  & \\
  \hline
 \multicolumn{7}{c}{HD~221170}  \\
  \hline
Mg & +0.11 & $-0.08$ & $-0.15$ &  +0.06 &  0.22 & 0.11 \\
Al & +0.07 & $-0.03$ & $-0.02$ &  +0.06 &  0.09 & 0.09 \\
Si & +0.08 & $ 0.01$ & $-0.10$ &  +0.02 &  0.13 & 0.09 \\
Fe & +0.14 & $-0.02$ & $-0.07$ &  +0.05 &  0.17 & ... \\
Sr & +0.02 & $-0.14$ & $-0.24$ &$-0.06$&   0.29 & 0.29 \\
  \hline
 \multicolumn{7}{c}{2MASS J0954+5246}  \\
  \hline
Na & +0.05 & $ 0.13$ & $-0.22$ & $-0.02$ &  0.26 & 0.25 \\
Mg & +0.09 & $-0.04$ & $-0.09$ &  +0.02 &  0.13 & 0.08 \\
Al & +0.07 & $-0.03$ & $-0.01$ &  +0.02 &  0.07 & 0.11 \\
Si & +0.07 & $ 0.06$ & $-0.06$ &  +0.01 &  0.10 & 0.09 \\
Fe & +0.16 & $-0.04$ & $-0.07$ &  +0.01 &  0.18 & ... \\
Sr & +0.05 & $ 0.13$ & $-0.22$ &$-0.02$&  0.26 & 0.25 \\
     \hline
    \end{tabular}}\label{tab:error}
\end{table}

\normalsize
\begin{longtable}{lrrrrr}
  \caption{Mg Abundances}\label{tab:mg}
  \hline\noalign{\vskip3pt}
  Object & $\log \epsilon$(Mg) & [Mg/Fe] & $N$ & $\sigma$ & Error$_{\rm total}$ \\
   \hline\noalign{\vskip3pt} 
\endfirsthead      
\endhead
\hline\noalign{\vskip3pt} 
\endfoot
HD~4306 & 5.19 & +0.61 & 3 & 0.18 & 0.13 \\
HD~6268         & 5.49 & +0.60 & 3 & 0.13 & 0.11 \\
HD~8724 & 6.19 & +0.43 & 2 & ... & 0.14 \\
HD~13979 & 5.20 & +0.46 & 3 & 0.18 & 0.11 \\
HD~85773         & 5.50 & +0.42 & 5 & 0.17 & 0.13 \\
HD~88609         & 5.17 & +0.67 & 4 & 0.12 & 0.12 \\
HD~107752         & 5.15 & +0.36 & 3 & 0.03 & 0.11 \\
HD~108577         & 5.79 & +0.42 & 4 & 0.15 & 0.14 \\
HD~110184         & 5.53 & +0.56 & 5 & 0.09 & 0.12 \\
HD~115444         & 5.16 & +0.55 & 4 & 0.05 & 0.10 \\
HD~118055     & 6.13 & +0.40 & 5 & 0.12 & 0.12 \\
HD~122563         & 5.30 & +0.49 & 6 & 0.09 & 0.09 \\
HD~187111         & 6.64 & +0.64 & 2 & ... & 0.16 \\
HD~204543         & 6.17 & +0.50 & 5 & 0.09 & 0.12 \\
HD~221170 & 5.88 & +0.41 & 6 & 0.06 & 0.11 \\
HD~237846         & 5.09 & +0.30 & 7 & 0.11 & 0.09 \\
BD$+03^{\circ}2782$          & 5.90 & +0.33 & 8 & 0.18 & 0.13 \\
BD$+30^{\circ}2611$          & 6.38 & +0.26 & 7 & 0.09 & 0.11 \\
BD$+44^{\circ}493$ & 4.51 & +0.74 & 3 & 0.09 & 0.10 \\
BD$+80^{\circ}245$          & 5.39 & $-$0.19 & 6 & 0.11 & 0.12 \\
BD$-02^{\circ}5957$ & 5.06 & +0.51 & 7 & 0.10 & 0.09 \\
BD$-07^{\circ}2674$       & 5.01 & +0.51 & 6 & 0.16 & 0.10 \\
BD$-14^{\circ}5890$        & 5.96 & +0.52 & 5 & 0.16 & 0.13 \\
BD$-15^{\circ}5781$          & 5.07 & +0.23 & 3 & 0.12 & 0.12 \\
BD$-18^{\circ}271$          & 5.53 & +0.57 & 6 & 0.14 & 0.12 \\
BD$-18^{\circ}5550$          & 5.15 & +0.58 & 3 & 0.13 & 0.11 \\
BD$-20^{\circ}6008$          & 5.38 & +0.64 & 3 & 0.23 & 0.13 \\
CS~29502--092          & 5.01 & +0.28 & 5 & 0.03 & 0.08 \\
CS~30314--067         & 4.98 & +0.35 & 4 & 0.09 & 0.12 \\
CS~31082--001          & 5.27 & +0.55 & 5 & 0.12 & 0.10 \\
HE~1116--0634         & 4.70 & +0.76 & 3 & 0.15 & 0.10 \\
HE~1320--1339         & 5.01 & +0.45 & 4 & 0.06 & 0.11 \\
HE~1523--0901         & 5.06 & +0.25 & 6 & 0.16 & 0.10 \\
LAMOST J0032+4107 & 4.99 & +0.01 & 2 & ... & 0.12 \\
LAMOST J0040+2729 & 5.29 & +0.39 & 4 & 0.15 & 0.09 \\
LAMOST J0748+4613 & 4.92 & +0.34 & 8 & 0.16 & 0.10 \\
LAMOST J2109+1725 & 5.56 & +0.46 & 4 & 0.11 & 0.10 \\
LAMOST J2114$-$0616 & 5.49 & +0.37 & 6 & 0.08 & 0.12 \\
LAMOST J2217+2104 & 5.00 & +1.33 & 7 & 0.12 & 0.09 \\
LAMOST J2347+2851 & 5.69 & +0.36 & 4 & 0.05 & 0.13 \\
2MASS J0643+5934 & 5.43 & +0.73 & 3 & 0.19 & 0.13 \\
2MASS J0954+5246 & 5.22 & +0.57 & 8 & 0.10 & 0.09 \\
2MASS J2016$-$0507 & 5.12 & +0.33 & 6 & 0.07 & 0.08 \\
2MASS J2145+1249 & 5.10 & +0.60 & 6 & 0.10 & 0.09 \\
\end{longtable}

\begin{longtable}{lrrrrr}
  \caption{Si Abundances}\label{tab:si}
  \hline\noalign{\vskip3pt}
  Object & $\log \epsilon$(Si) & [Si/Fe] & $N$ & $\sigma$ & Error$_{\rm total}$ \\
   \hline\noalign{\vskip3pt} 
\endfirsthead      
\endhead
\hline\noalign{\vskip3pt} 
\endfoot
HD~4306    & 5.25 & +0.76 & 19 & 0.11 & 0.10 \\
HD~6268    & 5.48 & +0.68 & 25 & 0.13 & 0.10 \\
HD~8724    & 6.16 & +0.49 & 29 & 0.13 & 0.09 \\
HD~13979   & 5.40 & +0.75 & 21 & 0.14 & 0.10 \\
HD~85773   & 5.53 & +0.54 & 26 & 0.20 & 0.10 \\
HD~88609   & 5.16 & +0.74 & 14 & 0.19 & 0.11 \\
HD~107752  & 5.14 & +0.44 & 16 & 0.14 & 0.10 \\
HD~108577  & 5.86 & +0.58 & 27 & 0.19 & 0.10 \\
HD~110184  & 5.51 & +0.63 & 27 & 0.16 & 0.09 \\
HD~115444  & 5.05 & +0.53 & 14 & 0.13 & 0.10 \\
HD~118055  & 6.06 & +0.43 & 29 & 0.13 & 0.09 \\
HD~122563  & 5.29 & +0.57 & 24 & 0.11 & 0.10 \\
HD~187111  & 6.36 & +0.45 & 24 & 0.17 & 0.09 \\
HD~204543  & 6.16 & +0.58 & 27 & 0.17 & 0.09 \\
HD~221170  & 5.82 & +0.44 & 29 & 0.14 & 0.09 \\
HD~237846  & 5.10 & +0.40 & 16 & 0.14 & 0.10 \\
BD$+03^{\circ}2782$ & 5.97 & +0.49 & 27 & 0.17 & 0.09 \\
BD$+30^{\circ}2611$ & 6.57 & +0.54 & 25 & 0.19 & 0.10 \\
BD$+44^{\circ}493$  & 4.18 & +0.50 & 5 & 0.12 & 0.11 \\
BD$+80^{\circ}245$ & 5.17 & $-0.32$ & 12 & 0.07 & 0.09 \\
BD$-02^{\circ}5957$ & 5.00 & +0.54 & 14 & 0.11 & 0.10 \\
BD$-07^{\circ}2674$ & 5.06 & +0.65 & 14 & 0.10 & 0.10 \\
BD$-14^{\circ}5890$ & 5.85 & +0.50 & 26 & 0.18 & 0.09 \\
BD$-15^{\circ}5781$ & 5.26 & +0.51 & 15 & 0.15 & 0.10 \\
BD$-18^{\circ}271$  & 5.44 & +0.57 & 26 & 0.14 & 0.09 \\
BD$-18^{\circ}5550$ & 5.11 & +0.63 & 18 & 0.13 & 0.10 \\
BD$-20^{\circ}6008$ & 5.26 & +0.61 & 20 & 0.18 & 0.10 \\
CS~29502--092 & 4.93 & +0.29 & 11 & 0.08 & 0.10 \\
CS~30314--067 & 5.04 & +0.50 & 15 & 0.18 & 0.10 \\
CS~31082--001 & 5.30 & +0.67 & 18 & 0.11 & 0.10 \\
HE~1116--0634 & 4.52 & +0.66 & 12 & 0.10 & 0.10 \\
HE~1320--1339 & 5.20 & +0.73 & 19 & 0.14 & 0.10 \\
HE~1523--0901 & 5.11 & +0.39 & 16 & 0.14 & 0.10 \\
LAMOST J0032+4107 & 5.29 & +0.40 & 9 & 0.13 & 0.10 \\
LAMOST J0040+2729 & 5.34 & +0.53 & 24 & 0.10 & 0.10 \\
LAMOST J0748+4613 & 5.00 & +0.52 & 15 & 0.07 & 0.09 \\
LAMOST J2109+1725 & 5.51 & +0.50 & 19 & 0.12 & 0.10 \\
LAMOST J2114$-$0616 & 5.65 & +0.62 & 22 & 0.18 & 0.10 \\
LAMOST J2217+2107 & 4.33 & +0.75 & 12 & 0.12 & 0.10 \\
LAMOST J2347+2851 & 5.71 & +0.47 & 24 & 0.15 & 0.09 \\
2MASS J0643+5934  & 5.55 & +0.94 & 12 & 0.21 & 0.11 \\
2MASS J0954+5246 & 5.18 & +0.61 & 21 & 0.11 & 0.10 \\
2MASS J2016$-$0507  & 5.03 & +0.33 & 15 & 0.12 & 0.10 \\
2MASS J2145+1249  & 4.95 & +0.54 & 15 & 0.09 & 0.10 \\
\end{longtable}

\begin{longtable}{lrrrr}
  \caption{Na Abundances}\label{tab:na}
  \hline\noalign{\vskip3pt}
  Object & $\log \epsilon$(Na) & [Na/Fe] & $N$ & Error$_{\rm total}$ \\
   \hline\noalign{\vskip3pt} 
\endfirsthead      
\endhead
\hline\noalign{\vskip3pt} 
\endfoot
HD~8724 & 4.49 & +0.09 & 1 &  0.18 \\
HD~85773 & 3.89 & +0.17 & 1 & 0.23 \\
HD~108577 & 4.19 & +0.18 & 1 &  0.22 \\
HD~110184 & 3.80 & +0.42 & 2 &  0.17 \\
HD~118055 & 4.44 & +0.08 & 2 &  0.15 \\
HD~122563 & 3.59 & +0.14 & 1 &  0.15 \\
HD~187111 & 4.67 & +0.03 & 1 &  0.21 \\
HD~204543 & 4.41 & +0.10 & 1 &  0.21 \\
HD~221170 & 3.86 & $-0.26$ & 1 & 0.19 \\
BD+30 2611  & 4.73 & $-0.03$ & 1 & 0.22 \\
BD-14 5890  & 4.46 & +0.38 & 1 & 0.22 \\
2MASS J0954+5246 &  3.62 & +0.32 & 1 & 0.14 \\
\end{longtable}

\begin{longtable}{lrrrrr}
  \caption{Al abundances}\label{tab:al}
  \hline\noalign{\vskip3pt}
  Object & $\log \epsilon$(Al) & [Al/Fe] & $N$ & $\sigma$ & Error$_{\rm total}$ \\
   \hline\noalign{\vskip3pt} 
\endfirsthead      
\endhead
\hline\noalign{\vskip3pt} 
\endfoot
HD~4306   & 3.30 & $-0.13$ & 1 & ...  & 0.16 \\
HD~6268   & 3.79 &  +0.05 & 2 & ... & 0.14 \\
HD~8724   & 4.66 &  +0.05 & 3 & 0.11 & 0.12 \\
HD~13979  & 3.73 &  +0.14 & 2 & ... & 0.15 \\
HD~85773  & 4.01 &  +0.08 & 4 & 0.09 & 0.14 \\
HD~88609  & 3.61 & +0.26 & 1 & ...  & 0.22 \\
HD~108577 & 3.91 & $-0.31$ & 1 & ...  & 0.21 \\
HD~110184 & 3.99 & +0.17 & 3 & 0.14 & 0.13 \\
HD~118055 & 4.62 & +0.04 & 3 & 0.09 & 0.12 \\
HD~122563 & 3.92 & +0.26 & 3 & 0.14 & 0.13 \\
HD~187111 & 5.26 & +0.41 & 4 & 0.21 & 0.12 \\
HD~204543 & 4.82 & +0.30 & 4 & 0.20 & 0.12 \\
HD~221170 & 4.34 & +0.01 & 3 & 0.08 & 0.12 \\
BD$+03^{\circ}2782$ & 4.59 & +0.17 & 5 & 0.18 & 0.12 \\
BD$+30^{\circ}2611$ & 5.07 & +0.10 & 6 & 0.28 & 0.12 \\
BD$-02^{\circ}5957$ & 3.64 & +0.24 & 1 & ...  & 0.16 \\
BD$-14^{\circ}5890$ & 4.22 & $-0.07$ & 4 & 0.12 & 0.13 \\
BD$-18^{\circ}271$  & 4.14 & +0.33 & 2 & ... & 0.13 \\
CS~31082--001 & 3.42 & $-0.15$ & 2 & ... & 0.14 \\
HE~1523--0901 & 3.60 & $-0.07$ & 1 & ...  & 0.18 \\
LAMOST J0032+4107 & ...  & ...  & 0 & ...  & ...  \\
LAMOST J0040+2729 & 3.78 & +0.02 & 3 & 0.16 & 0.13 \\
LAMOST J2109+1725 & 3.64 & $-0.31$ & 2 & ... & 0.14 \\
LAMOST J2114$-$0616 & 3.76 & $-0.21$ & 3 & 0.12 & 0.14 \\
LAMOST J2347+2851 & 4.17 & $-0.01$ & 4 & 0.10 & 0.12 \\
2MASS J0954+5246 & 3.61 & +0.10 & 3 & 0.19 & 0.13 \\
2MASS J2016$-$0507  & 3.53 & $-0.11$ & 1 & ...  & 0.16 \\
2MASS J2145+1249  & 3.81 & +0.46 & 2 & ... & 0.13 \\
\end{longtable}

\begin{longtable}{lrrrrr}
  \caption{Sr abundances}\label{tab:sr}
  \hline\noalign{\vskip3pt}
  Object & $\log \epsilon$(Sr) & [Sr/Fe] & $N$ &  $\sigma$ &  Error$_{\rm total}$ \\
\hline\noalign{\vskip3pt} 
\endfirsthead      
\endhead
\hline\noalign{\vskip3pt} 
\endfoot
HD~4306   & 0.35 & +0.50 & 3 & 0.09 & 0.26 \\
HD~6268   & 1.17 & +1.01 & 3 & 0.20 & 0.26 \\
HD~8724   & 1.79 & +0.76 & 3 & 0.14 & 0.30 \\
HD~13979  & 0.59 & +0.58 & 2 & ... & 0.27 \\
HD~85773  & 1.16 & +0.81 & 3 & 0.26 & 0.31 \\
HD~88609  & 0.70 & +0.92 & 3 & 0.10 & 0.28 \\
HD~107752 & 0.77 & +0.71 & 2 & ... & 0.27 \\
HD~108577 & 1.41 & +0.77 & 3 & 0.27 & 0.31 \\
HD~110184 & 1.36 & +1.12 & 3 & 0.19 & 0.30 \\
HD~115444 & 0.60 & +0.72 & 3 & 0.09 & 0.26 \\
HD~118055 & 1.68 & +0.68 & 3 & 0.12 & 0.30 \\
HD~122563 & 0.72 & +0.64 & 3 & 0.14 & 0.26 \\
HD~187111 & 1.94 & +0.67 & 3 & 0.17 & 0.30 \\
HD~204543 & 1.83 & +0.89 & 3 & 0.17 & 0.30 \\
HD~221170 & 1.41 & +0.66 & 3 & 0.14 & 0.30 \\
HD~237846 & 0.40 & +0.34 & 3 & 0.08 & 0.27 \\
BD+03 2782 & 1.88 & +1.04 & 3 & 0.13 & 0.30 \\
BD+30 2611 & 2.35 & +0.96 & 3 & 0.23 & 0.31 \\
BD+80 245  & 0.12 & $-0.73$ & 1 & ...  & 0.30 \\
BD-02 5957 & 0.67 & +0.85 & 3 & 0.14 & 0.26 \\
BD-07 2674 & 0.36 & +0.59 & 3 & 0.09 & 0.26 \\
BD-14 5890 & 1.40 & +0.69 & 3 & 0.15 & 0.31 \\
BD-15 5781 & 0.83 & +0.72 & 3 & 0.22 & 0.27 \\
BD-18 271  & 0.98 & +0.75 & 3 & 0.16 & 0.30 \\
BD-18 5550 & -0.51 & $-0.35$ & 2 & ... & 0.27 \\
BD-20 6008 & 0.56 & +0.55 & 3 & 0.24 & 0.27 \\
CS~29502--092 & 0.07 & +0.07 & 2 & ... & 0.26 \\
CS~30314--067 & 0.30 & +0.40 & 3 & 0.06 & 0.27 \\
CS~31082--001 & 1.32 & +1.33 & 3 & 0.15 & 0.26 \\
HE~1320--1339 & 0.94 & +1.11 & 3 & 0.21 & 0.27 \\
HE~1523--0901 & 1.20 & +1.12 & 3 & 0.12 & 0.27 \\
LAMOST J0032+4107 & 1.07 & +0.82 & 1 & ...  & 0.28 \\
LAMOST J0040+2729 & 1.10 & +0.93 & 3 & 0.12 & 0.26 \\
LAMOST J0748+4613 & 0.23 & +0.38 & 3 & 0.07 & 0.26 \\
LAMOST J2109+1725 & 0.50 & +0.13 & 2 & ... & 0.27 \\
LAMOST J2114$-$0616 & 1.49 & +1.10 & 3 & 0.22 & 0.31 \\
LAMOST J2347+2851 & 1.45 & +0.85 & 3 & 0.19 & 0.30 \\
2MASS J0643+5934 & 1.37 & +1.40 & 3 & 0.12 & 0.27 \\
2MASS J0954+5246 & 1.00 & +1.07 & 3 & 0.20 & 0.26 \\
2MASS J2145+1249 & 0.32 & +0.55 & 3 & 0.07 & 0.26 \\
\end{longtable}

\begin{table}
  \tbl{Average and standard deviation of abundance ratios}{%
    \begin{tabular}{lccccccccccc}
      \hline
 &  \multicolumn{5}{c}{$-3.5<$[Fe/H]$<-2.5$} && \multicolumn{5}{c}{$-2.5<$[Fe/H]$<-1.5$} \\     
\cline{2-6} \cline{8-12}
      &  \multicolumn{2}{c}{This work}  &&  \multicolumn{2}{c}{Literature} && \multicolumn{2}{c}{This work}  &&  \multicolumn{2}{c}{Literature} \\
\cline{2-3} \cline{5-6} \cline{8-9} \cline{11-12}
  & average & $\sigma$ && average & $\sigma$ && average & $\sigma$ && average & $\sigma$ \\
\hline
 $[$Mg/Fe$]$ &  0.46 & 0.15 && 0.38    & 0.19 && 0.43 & 0.10 && 0.27    &  0.18 \\
 $[$Al/Fe$]$ &  0.09 & 0.17 && $-$0.62 & 0.28 && 0.04 & 0.25 && $-$0.80 &  0.24 \\
 $[$Si/Fe$]$ &  0.58 & 0.14 && 0.54    & 0.28 && 0.51 & 0.05 && 0.44    &  0.19 \\
 $[$Sr/Fe$]$ &  0.68 & 0.47 && $-$0.26 & 0.58 && 0.74 & 0.19 && $-$0.14 &  0.43 \\
     \hline
  \end{tabular}}\label{tab:ave}
\end{table}

\clearpage




\end{document}